\documentclass[%
 reprint,
 amsmath,amssymb,
 aps,
]{revtex4-2}

\usepackage{graphicx}
\usepackage{dcolumn}
\usepackage{bm}
\usepackage{hyperref}
\usepackage{xcolor}

\begin{document}


\title{Acoustic wave tunneling across a vacuum gap between two piezoelectric crystals with arbitrary symmetry and orientation}

\author{Zhuoran Geng}
 \email{zhgeng@jyu.fi}
\author{Ilari J. Maasilta}%
\email{maasilta@jyu.fi}
\affiliation{%
Nanoscience Center, Department of Physics, University of Jyvaskyla, P. O. Box 35, FIN-40014 Jyv{\"a}skyl{\"a}, Finland
}%
\date{\today}

\begin{abstract}
It is not widely appreciated that an acoustic wave can "jump" or "tunnel" across a vacuum gap between two piezoelectric solids, nor has the general case been formulated or studied in detail. Here, we remedy that situation, by presenting a general formalism and approach to study such an acoustic tunneling effect between two arbitrarily oriented anisotropic piezoelectric semi-infinite crystals. The approach allows one to solve for the reflection and transmission coefficients of all the partial wave modes, and is amenable to practical numerical or even analytical implementation, as we demonstrate by a few chosen examples. 
The formalism can be used in the future for quantitative studies of the tunneling effect in connection not only with the manipulation of acoustic waves, but with many other areas of physics of vibrations such as heat transport, for example.     
\end{abstract}

\maketitle

\section{Introduction}\label{sec:introduction}
Acoustic waves in solids (also known as elastic waves) have many applications ranging from acoustic wave filters for mobile phones, mechanical resonators for sensors, acousto-optical modulators for optical signal processing, to ultrasonic imaging devices, to name a few \cite{RoyerII}. They are often generated with the help of piezoelectric (PE) transducers, converting an electrical signal to an acoustic wave, as piezoelectricity couples acoustic deformations and electric fields \cite{RoyerI,Auld1973}. This also means that the elastic waves in piezoelectric materials are not purely elastic, but contain electric waves as a by-product. To be more descriptive, such waves are sometimes also called acoustoelectric or electroacoustic waves. The main very well understood effect of piezoelectricity on the propagation of acoustic waves is that the acoustic velocities are slightly modified, due to the piezoelectric "stiffening" of the effective elastic constants for wave propagation \cite{RoyerI,Auld1973}.    

However, when considering wave transmission and reflection problems, another intriguing and much less widely known effect due to piezoelectricity can happen: a bulk elastic wave can be transmitted across a vacuum gap between two piezoelectric solids. This transmission is not possible for purely elastic waves, which by definition cannot exist in vacuum, but is made possible by the evanescent electric field components of the electroacoustic waves extending into the vacuum gap. The effect works for gap sizes of the order of the acoustic wavelength, which is much longer than the length scale of other possible  mechanisms that can couple bulk acoustic wave energy across vacuum gaps in the nanometer to sub-nanometer scale (such as van der Waals, Casimir and electrostatic interactions discussed in the context of heat transfer, see \cite{Budaev2011,Ezzahri2014,Xiong2014,Chiloyan2015,Pendry2016,Volokitin2020}). As such, the phenomenon is highly analogous to quantum mechanical tunneling of a particle through a classically forbidden region, and for this reason, we also call this effect  "acoustic wave tunneling" or "phonon tunneling", terminology that was already introduced by others \cite{Balakirev1978,Prunnila2010,Chiloyan2015,Pendry2016,Volokitin2020}.       

To our knowledge, acoustic wave tunneling mediated by piezoelectricity was first discussed theoretically by Kaliski\cite{Kaliski1966} for the case of horizontally polarized shear (SH) waves in a cubic piezoelectric crystal in the limit of zero gap width. Later, in an important seminal work Balakirev and Gorchakov \cite{Balakirev1977} extended the calculations for the same SH wave mode for finite gap widths (with the cubic axes aligned with the surfaces). They also provided results for hexagonal crystals with the $c_6$-symmetry axis oriented parallel to the surfaces, still considering only the SH wave mode, and plotted examples for the transmission coefficient vs. incident angle for Bi$_{12}$GeO$_{20}$ (cubic) and LiIO$_3$ (hexagonal). An important result of that study was that the transmission coefficient was shown to be large and even approaching unity for angles close to glancing incidence. An experimental study by the same authors \cite{Balakirev1978} with ultrasound ($f = 15$ MHz) using LiIO$_3$ crystals confirmed the phenomenon with observed transmission coefficients up to $\sim 0.5$.

These early studies used the standard piezoelectrically stiffened elasticity theory \cite{Auld1973,Auld1981} and were focused on finding explicit solutions, available only for the highest symmetry crystal orientations and for the simplest wave modes, therefore providing only expressions with no generality.

Much more recently, Prunnila and Meltaus revisited the topic in the context of thermal transport using a scattering matrix approach \cite{Prunnila2010}, and provided results for energy transmission coefficients as a function of the angle of incidence and wave vector. However, their approach assumed isotropic properties of the materials, a simplified single component piezoelectric tensor, no PE stiffening,  and the results were limited to a single symmetry direction of the "crystal". Within these approximations only two modes contribute.   

On the other hand, to study anisotropic piezoelectric insulators more generally, Barnett and Lothe \cite{Barnett1975,Lothe1976} extended the so called sextic Stroh formalism, an elegant and mathematically powerful tool to analyze anisotropic elasticity\cite{Stroh1962,Chadwick1977,Ting1996book,Ting2000}, to an eight-dimensional framework for {\em arbitrary} anisotropic piezoelectric crystals. This extended Stroh formalism was further developed by several authors \cite{Alshits1989,Alshits1990,Alshits1991,Chung1995,Akamatsu1997,Hwu2008}and has been successfully applied from the analysis of reflection of bulk electroacoustic waves \cite{Alshits1989,Alshits1990,Alshits1991} and anisotropic piezoelectric surface acoustic waves (SAW)\cite{Lothe1976,Lyubimov1980,Darinskii2003} to the gap waves (GW) \cite{ALSHITS1993,Alshits1994,Darinskii2006}, which are surface waves guided and coupled by a gap between two piezoelectric surfaces\cite{Gulyaev1976,Gulyaev1977}.  It has also been used in material science applications for piezoelectric ceramics and composites \cite{Pak1992,Liang1995,Lu2006} and recently\cite{Darinskii2019} also to study the control SAW propagation using piezoelectric phononic crystals\cite{Benchabane2006}. 

Even though the framework of extended piezoelectric Stroh formalism was developed some time ago, only a limited number of investigations have been carried out to study the phenomenon of bulk acoustic wave tunneling. Al'shits \textit{et.al.},\cite{ALSHITS1993} introduced formally a general solution of reflection and transmission coefficient for an incident slow quasi-transverse bulk wave. 
Later, Darinskii developed this framework further\cite{Darinskii1997,Darinskii1998b}  and investigated the reflection and transmission mediated by
  the leaky gap wave \cite{Darinskii2006}. In these studies, only single transmitted bulk wave mode was considered, and the resonance conditions of the leaky gap waves were usually applied.

The purpose of this work is 
to demonstrate a general formalism and solution for transmission of elastic waves across a vacuum gap that is applicable to any incident bulk wave mode in any anisotropic crystallographic orientation. Furthermore, an alternative approach to the direct solution will be presented. In this method, the scattering problems of semi-infinite piezoelectric half-spaces are solved independently for both crystals using the extended Stroh formalism, and the reflection and transmission coefficients of the tunneling are acquired with a simple factor, which is determined from multiple reflections of evanescent electric waves inside the vacuum gap. To our knowledge, such an interpretation of acoustic wave tunneling has not been discussed in literature, although the multiple reflection picture has been widely adopted in the field of near-field electromagnetic wave tunneling\cite{Polder1971,Pendry1999,Joulain2005}.

This work is organized as follows: We first briefly introduce the the main aspects of the extended Stroh formalism for plane interface scattering problems using generally applicable coordinate setup and plane wave functions in Section.\ref{sec:stroh_formalism}. The tunneling problem for the plane-plane geometry is then solved in Section \ref{sec:tunneling}, first by directly applying the boundary conditions to the Stroh eigenfunctions, then followed by the alternative approach of using multiple reflection factor. In Section.\ref{sec:example}, we then present a few illustrative examples: first an analytical solution for a hexagonal crystal with a high symmetry orientation derived using both methods, and finally numerical calculations of tunneling transmission coefficients for a couple of different crystallographic cuts of a hexagonal crystal in different orientations. At the end in section \ref{sec:conclusions}, we present conclusions and outlook on the applications of this study. 

\section{Extended Stroh formalism for scattering problems}\label{sec:stroh_formalism}
\begin{figure}[h]
	\centering
	\includegraphics[width=1\linewidth]{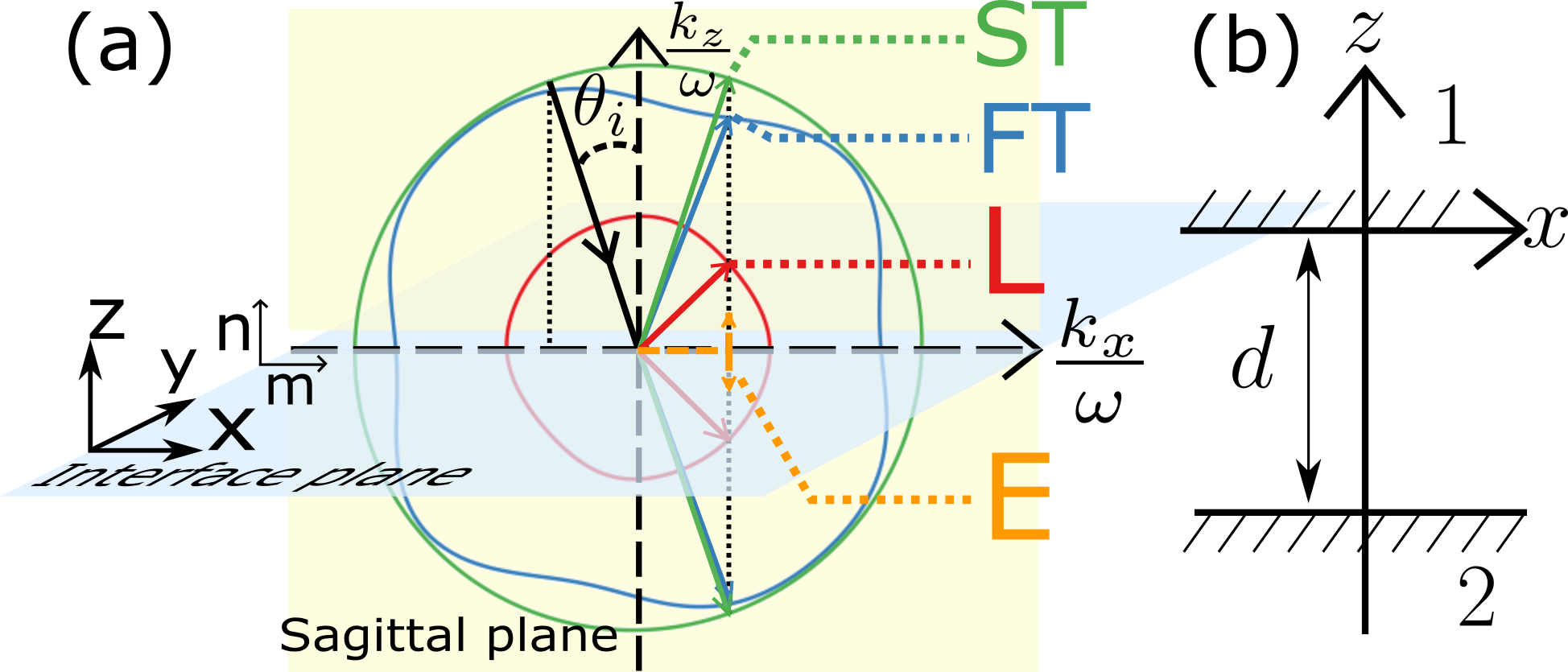}
	\caption{(a) Schematic of the single interface scattering problem with the coordinate system used, with the interface plane  at $z=0$ and the sagittal plane at $y=0$.  The scattering and mode conversions of an incident bulk slow transverse (ST) wave from the angle $\theta_i$ are shown, with a hypothetical set of slowness surfaces. $\pmb{m}$ and $\pmb{n}$ are the unit vectors of the $x-$axis and $z-$axis. Under the quasistatic approximation, four wave modes are shown: longitudinal (L), fast transverse (FT), slow transverse (ST), and quasistatic electric potential (E). (b) Two piezoelectric media $1$, $2$ are separated by a vacuum gap of width $d$. }
	\label{fig:theoryconfig}
\end{figure}

We consider an incident acoustic plane wave with a wave vector $\pmb{k}$ in an anisotopic piezoelectric medium using Cartesian coordinates $\pmb{r}=[x,y,z]^T$ ($[...]^T$ stands for transposition), with an interface plane $\pmb{n}\cdot\pmb{r}=z=0$ between two media, and a plane of incidence (sagittal plane) $(\pmb{n}\times\pmb{m})\cdot\pmb{r}=y=0$, where $\pmb{n}$ is the unit normal vector of the interface plane and  $\pmb{m}$ the unit vector parallel to the interface and sagittal planes. With our coordinate system, they are the unit vectors of the $z-$axis and $x-$axis, respectively [see Fig.\ref{fig:theoryconfig}(a)]. Without losing generality, we always consider that the incident bulk wave has a positive $x$-component of the wave vector $k_x > 0$, and propagates in the sagittal plane (the wave vector is contained in the plane), but the sagittal plane has a rotational degree of freedom with respect to the normal of interface plane ($z$-axis)\footnote{The sagittal plane has a rotational degree of freedom with respect to the normal of the interface plane (azimuth angle), which is equivalent to the rotation of the crystal azimuth angle $\varphi$. For the sake of simplicity and to avoid the duplication of the effect of this degree of freedom, we unambiguously take into account the azimuth angle by the rotation of the crystal (see Appendix \ref{apd:crystal_rotation}).}.  The piezoelectric medium is characterized by its density $\rho$, piezoelectric stress tensor $e_{iL}$, elastic stiffness tensor at constant electric field $c_{KL}^E$ and electric permittivity tensor at constant strain $\epsilon_{ij}^S$, where $i,j=x,y,z$ are the Cartesian coordinate indices and $K,L=1,...,6$ are the abbreviated Voigt indices. 

The sound velocities $v=\omega/k$ are typically more than four orders of magnitude slower than the speed of light, therefore it is possible and customary to apply the quasistatic approximation\cite{Auld1973} to the piezoelectric scattering problems, ignoring the magnetic field. Under such conditions, the propagation of a time-harmonic plane wave $\sim\exp(-i\pmb{k}\cdot\pmb{r}+i\omega t)$ with wave vector $\pmb{k}$ and angular frequency $\omega$ is governed by the elastic equation of motion $\nabla\cdot\pmb{\sigma} = \rho\partial^2\pmb{u}/\partial t^2$ and just one of the Maxwell's equations (Gauss's law) $\nabla\cdot\pmb{D}=0$, which together with the piezoelectric constitutive relations read in the matrix notation \cite{Auld1973}:
\begin{equation}\label{eq:const_equs}
	\begin{aligned}
		ik_{iK}\sigma_{K} &= \rho\omega^2u_i \\
		-ik_iD_{i} &= 0 \\
		\sigma_{K} &= -ic^E_{KL}k_{Lj}u_{j}-ie_{Kj}k_j\varPhi \\
		D_i &= -ie_{iL}k_{Lj}u_j + i\epsilon^S_{ij}k_j\varPhi, \\
	\end{aligned}
\end{equation}
where $\pmb{\sigma}$, $\pmb{u}$, $\pmb{D}$ are the elastic stress, mechanical displacement and electric displacement fields, respectively, $\varPhi$ is the electric potential, and $k_{iK}$ is a $3\times6$ matrix defined by the wave vector components \cite{Auld1973} (see Appendix \ref{apd:voigtnotation} for more details). As usual, repeated indices are summed. 

An incident plane wave is scattered into a linear combination of partial waves at the interface, which are either reflected or transmitted, and can also be inhomogeneous modes (An example where three reflected and transmitted homogeneous bulk acoustic waves and an inhomogeneous electric potential wave are generated is depicted in Fig.\ref{fig:theoryconfig}(a).) The general solutions of such partial waves that satisfy the governing equations can be written \cite{Alshits1989,Alshits1990,Alshits1991}as:
\begin{equation}
	\begin{aligned}
		\pmb{u} & = \sum_{\alpha}b_\alpha\pmb{A}_\alpha
		e^{-ik_x(x+p_\alpha z)+i\omega t}
		\\
		\varPhi & = \sum_{\alpha}b_\alpha\phi_\alpha
		e^{-ik_x(x+p_\alpha z)+i\omega t}
		\\
		\pmb{n}\cdot\pmb{\sigma} & = ik_x\sum_{\alpha}b_\alpha\pmb{L}_\alpha
		e^{-ik_x(x+p_\alpha z)+i\omega t}
		\\
		\pmb{n}\cdot\pmb{D} & = ik_x\sum_{\alpha}b_\alpha D_\alpha^n
		e^{-ik_x(x+p_\alpha z)+i\omega t} \ ,
		\\
	\end{aligned} 
	\label{eq:stroh_field_functions}
\end{equation}
in which $\pmb{A}_\alpha,\phi_\alpha,\pmb{L}_\alpha,D_\alpha^n$ are normalized constants describing the displacement (polarization vector), the electric potential, the traction force and the normal projection of the electric displacement of a partial wave mode $\alpha$, respectively. $b_\alpha$ is the dimensionless amplitude of the partial wave, and $p\equiv k_z/k_x$ where $k_z$ and $k_x$ are the normal and parallel components of the $\pmb{k}-$vector. To avoid redundant writing in expressions, we omit from now on the common phase factor $\exp(-ik_xx+i\omega t)$ shared by all solutions. 


With the above partial wave formulation, the solution of the governing equations, Eqs. (\ref{eq:const_equs}), reduces to determining the eigenvalues $p_{\alpha}$ and eigenvectors $\pmb{\xi}_\alpha$ of an eight-dimensional eigenvalue problem \cite{Barnett1975,Lothe1976} with a $8\times8$ real matrix $\pmb{N}$: 
\begin{equation}
	\pmb{N}(v_x)\pmb{\xi}_\alpha = p_\alpha\pmb{\xi}_\alpha \ ,
	\label{eq:stroh_eigenfunc}
\end{equation}
where the matrix $\pmb{N}(v_x)$ depends on the phase velocity along the interface $v_x\equiv\omega/k_x$, a conserved quantity due to continuity conditions on the boundary, and the orientation and material of the crystal. The eight-component eigenvector for mode $\alpha$ is defined as $\pmb{\xi}_\alpha=[\pmb{A}_\alpha,\phi_\alpha,\pmb{L}_\alpha,D_\alpha^n]^T$. The derivation of Eq.\eqref{eq:stroh_eigenfunc} with the detailed definition of $\pmb{N}(v_x)$ is presented in Appendix \ref{apd:stroh_formalism}.

The matrix $\pmb{N}$ also satisfies the symmetry relation $(\hat{\pmb{T}}\pmb{N})^T=\hat{\pmb{T}}\pmb{N}$, 
where the $8\times8$ matrix $\hat{\pmb{T}}$ is given by
\begin{equation*}
\begin{aligned}
	&\hat{\pmb{T}} = 
	&\begin{bmatrix}
		\hat{\pmb{0}} & \hat{\pmb{I}} \\
	 \hat{\pmb{I}} & \hat{\pmb{0}}
	\end{bmatrix},
\end{aligned}
\end{equation*}
with $\hat{\pmb{I}}$ and $\hat{\pmb{0}}$ the $4\times4$ unit and zero matrices, respectively
\cite{Stroh1962,Malen1970,Chadwick1977}. This relation provides an orthonormalization condition:
\begin{equation}
	\pmb{\xi}_\alpha^T\hat{\pmb{T}}\pmb{\xi}_\beta 
	= \delta_{\alpha\beta},\ \alpha,\beta=1,...,8 \ ,
	\label{eq:stroh_inner_basis}
\end{equation}
where $\delta_{\alpha\beta}$ is the Kronecker delta, and ensures a unique and complete set of solutions for the extended Stroh eigenfunction \footnote{The above results are strictly true only if $\pmb{N}$ is non-degenerate (has a non-zero determinant). Slightly modified eigenvectors and normalization conditions have been determined in the opposite case \cite{Darinskii2003} taking place at the exact conditions for a critical angle (transonic state), where the reflected bulk wave carries energy only along the interface. Our discussion is meant for the general case to facilitate numerical computation, thus these conditions are special cases that do not have to be considered here, as numerical computation can be done very close to the exact conditions.}.

At this point it is good to point out that the above "Stroh-normalization", widely used in literature as it is, does not keep the physical units (as $\pmb{\xi}_\alpha^T\hat{\pmb{T}}\pmb{\xi}_\alpha$ has units of force), but introduces computationally useful "Stroh-units". This is not a problem, as the units cancel out in the end if transmission and reflection amplitudes are the observables to be calculated.
  
Totally eight partial wave mode solutions can be obtained from Eq.\eqref{eq:stroh_eigenfunc}, containing complex eigenvalues $p_\alpha=p_\alpha'+ip_\alpha''$ and the associated eigenvectors $\pmb{\xi}_\alpha$ with $\alpha=1,...,8$ (with $p_\alpha'$ denoting the real part and $p_\alpha''$ the imaginary part). These partial waves can be either homogeneous plane waves ($p_\alpha''=0$) or inhomogeneous waves ($p_\alpha''\neq0$). For an inhomogeneous wave mode, the scattering direction of the wave is determined by the imaginary part of the eigenvalue such that if $p_\alpha''>0$ ($p_\alpha''<0$) the wave is transmitted (reflected), to ensure decaying solutions at infinity. For a plane wave mode, the direction of the power flow normal to the interface is examined. By investigating the time-averaged acoustic Poynting vector component normal to the interface \cite{Alshits1989}
\begin{equation}\label{eq:Poynting_vector}
	P_{n\alpha}\equiv\pmb{n}\cdot\pmb{P}_{AV,\alpha} 
	= -\frac{\omega k_x}{4}|b_\alpha|^2\pmb{\xi}_\alpha^T\hat{\pmb{T}}\pmb{\xi}_\alpha^* \ ,
\end{equation} 
one can determine which wave mode is transmitted ($P_{n\alpha}<0$) or reflected ($P_{n\alpha}>0$). 

If Stroh-normalization is used, Eq.\eqref{eq:Poynting_vector} simplifies even further for the bulk modes. For them, the eigenvalues $p_\alpha$ are real, which means that the eigenvectors are real as well, $\pmb{\xi}_\alpha^* = \pmb\xi_\alpha$, as $\pmb{N}$ is real. We then have $\pmb{\xi}_\alpha^T\hat{\pmb{T}}\pmb{\xi}_\alpha^* = \pmb{\xi}_\alpha^T\hat{\pmb{T}}\pmb{\xi}_\alpha = 1$, and see from Eq.\eqref{eq:Poynting_vector} that the power transmission and reflection coefficients (ratios of Poynting vector normal components) are simply given by the ratio $|b_\alpha/b_{in}|^2$. This is one justification for the usefulness of the used Stroh-normalization. 

\section{Bulk acoustic wave tunneling across a vacuum gap}\label{sec:tunneling}
To formulate a generalized expression for bulk acoustic wave tunneling  across a vacuum gap between two adjacent piezoelectric solids, we consider a geometry which consists of two parallel semi-infinite piezoelectric half-spaces (medium 1 and medium 2) separated by a vacuum gap of distance $d$, as shown in Fig.\ref{fig:theoryconfig}(b). The incident wave is propagating towards the gap from the positive $z$-coordinate direction 
in the sagittal plane, and the two solid-vacuum interfaces are located at $\pmb{n}\cdot\pmb{r}=0$ and $\pmb{n}\cdot\pmb{r}=-d$.

For a given incident wave propagating in a given crystal orientation, 
the wave vector and phase velocity components along the interface ($k_x$ and $v_x$) are known. Therefore, the unknowns left in the partial wave solutions in Eqs.\eqref{eq:stroh_field_functions} are the eigenvectors $\pmb{\xi}_\alpha$ and eigenvalues $ p_\alpha$ of the Stroh eigen-equation Eq.\eqref{eq:stroh_eigenfunc}, as well as the amplitude factors $b_\alpha$. The eigenvectors and eigenvalues can readily be solved with the knowledge of the material, the crystallographic orientation and $v_x$, whereas the determination of $b_\alpha$ requires solving the boundary conditions of the solid-vacuum interfaces.

We assume for this study that both interfaces are mechanically free and without electrodes or net charge density on the surface, i.e. electrically free. For such a case, there are conditions for the continuity of the electric potential and the normal component of the electric displacement, giving for the boundary conditions
\begin{equation}
	\begin{aligned}
		\varPhi^{(i)} &=\varPhi_{V} \\
		\pmb{n}\cdot\pmb{D}^{(i)} &= 
		\pmb{n}\cdot\pmb{D}_{V} \\
		\pmb{n}\cdot\pmb{\sigma}^{(i)} &= \pmb{0}
	\end{aligned} \ ,
	\label{eq:general_boundcond}
\end{equation}
in which the superscript $i=1,2$ indicates the medium index and the subscript $V$ represents the fields in the vacuum gap.

In the vacuum region, the electric potential wave must satisfy the Laplace equation $\nabla^2\varPhi_V=0$, which for the plane waves leads to the condition $k_x^2+k_z^2 = 0$.  Thus, it can be expressed in terms of two partial wave modes with $k_z = \pm ik_x$. 
Following the form of the general solutions for the electric potential in Eq.\eqref{eq:stroh_field_functions} leads to a solution with decaying and increasing exponentials 
[omitting the common phase factor $\exp(-ik_xx+i\omega t)]$  
\begin{equation}
		\varPhi_V = b_{V_+}\phi_{V_+}e^{k_xz} + b_{V_-}\phi_{V_-}e^{-k_xz},
		\label{eq:vac_field_functions1}
\end{equation}		
and the normal component of the vacuum electric displacement can then be calculated directly from $\pmb{D}_V = -\epsilon_0\nabla\cdot\varPhi_V $, giving 
\begin{equation}		
		\pmb{n}\cdot\pmb{D}_V =
		-\epsilon_0k_xb_{V_+}\phi_{V_+}e^{k_xz}
		+\epsilon_0k_xb_{V_-}\phi_{V_-}e^{-k_xz}. 
	\label{eq:vac_field_functions2}
\end{equation}

From the form of solutions above, the analogy with quantum mechanical tunneling is apparent.


Comparing the result for the electric displacement in Eq.\eqref{eq:vac_field_functions2} with the definitions of the Stroh formalism, Eqs.\eqref{eq:stroh_field_functions}, we see that the components for the electric displacement $D_{V_\pm}$ and for the potential 
$\phi_{V_\pm}$ satisfy a simple relation $D_{V_\pm}=\pm i\epsilon_0\phi_{V_\pm}$. In addition, we have $\phi_{V_\pm}=1/\sqrt{\pm2i\epsilon_0}$, as can be readily calculated from the Stroh-normalization condition $2\phi_{V_\pm}D_{V_\pm}=1$ obtained from Eq.\eqref{eq:stroh_inner_basis} by setting the vacuum eigenvector components associated with the displacement and traction force to zero: $\pmb{A}_{V_\pm}=\pmb{0}$, $\pmb{L}_{V_\pm}=\pmb{0}$.

By inserting the general solutions of Eqs.\eqref{eq:stroh_field_functions}, Eq.\eqref{eq:vac_field_functions1} and Eq. \eqref{eq:vac_field_functions2} into the boundary conditions in Eqs.\eqref{eq:general_boundcond}, we obtain two sets of linear equations to express the boundary conditions at both interfaces as
\begin{equation}\label{eq:boundary_condition}
	\begin{aligned}			b_{in}^{(1)}\pmb{U}_{in}^{(1)}+\sum_{\alpha=1}^4b_{\alpha}^{(1)}\pmb{U}_{\alpha}^{(1)} &=
		b_{V_+}\pmb{U}_{V_+}+b_{V_-}\pmb{U}_{V_-} 
		\ , \\ 
		\sum_{\alpha=1}^4\tilde{b}_{\alpha}^{(2)}\pmb{U}_{\alpha}^{(2)} &= 
		b_{V_+}\pmb{U}_{V_+}e^{-k_xd} + b_{V_-}\pmb{U}_{V_-}e^{k_xd} \ ,
	\end{aligned}	
\end{equation}
in which we introduce a $5\times1$ column vector $\pmb{U}_\gamma^{(i)}=[\phi_\gamma^{(i)},D_\gamma^{n,(i)},\pmb{L}^{(i)}_\gamma]^T$ for the wave mode $\gamma=in,\alpha$, where the subscript $in$ indicates the incident wave mode, $\alpha=1,...,4$ corresponds to the four physically allowed wave modes in the corresponding media $i=1,2$ (the reflected and transmitted modes, respectively), and $\tilde{b}^{(2)}_\alpha \equiv b^{(2)}_\alpha\exp(ip^{(2)}_\alpha k_xd)$. As $\pmb{U}_\gamma^{(i)}$ are known and defined by the Stroh eigenvectors  (more explicit expressions can be found in Appendix \ref{apd:matrix_approach}), Eqs.\eqref{eq:boundary_condition} can be used to solve for the partial wave amplitudes $b_{\alpha}^{(i)}$, $b_{V_\pm}$, giving us finally the transmission and reflection amplitude coefficients $t^{(2)}_\alpha\equiv \tilde{b}^{(2)}_\alpha/b^{(1)}_{in}$ and $r^{(1)}_\alpha\equiv b^{(1)}_\alpha/b^{(1)}_{in}$. 


We have considered two approaches to solve Eqs.\eqref{eq:boundary_condition}: First, by directly solving the combined boundary conditions of both interfaces with matrix algebra, and second, by using a multiple reflection factor to connect the separate solutions on each interface. Both approaches are discussed in the following and give identical results. 

\subsection{Combined boundary conditions approach}\label{sec:comb_boundaries}
In the first approach, where the boundary conditions are solved directly, we introduce two $8\times5$ matrices $\pmb{M}_1$ and $\pmb{M}_2$:
\begin{equation}
	\pmb{M}_1 =
	\begin{bmatrix}
		\pmb{V}_1^{-1} & \hat{\pmb{O}}_{(2\times3)} \\
		\hat{\pmb{O}}_{(3\times2)} & \hat{\pmb{I}}_{(3)} \\
		\hat{\pmb{O}}_{(3\times2)} & \hat{\pmb{O}}_{(3\times3)} \\
	\end{bmatrix}, \
	\pmb{M}_2 =
	\begin{bmatrix}
		\pmb{V}_2^{-1} & \hat{\pmb{O}}_{(2\times3)} \\
		\hat{\pmb{O}}_{(3\times2)} & \hat{\pmb{O}}_{(3\times3)} \\
		\hat{\pmb{O}}_{(3\times2)} & \hat{\pmb{I}}_{(3)} \\
	\end{bmatrix} \ ,
\label{eq:vac_M}
\end{equation}
in which $\hat{\pmb{O}}_{(m\times n)}$ and $\hat{\pmb{I}}_{(m)}$ are the $m\times n$ zero matrix and the $m\times m$ identity (unit) matrix. $\pmb{V}_1,\pmb{V}_2$ are $2\times 2$ matrices 
depending only on the wave vector component $k_x$ along the interface and the size of vacuum gap $d$:
\begin{equation*}
	\pmb{V}_{1} =
	\begin{bmatrix}
		\phi_{V+} & \phi_{V-} \\
		D_{V+} & D_{V-} 
	\end{bmatrix},\ 
	\pmb{V}_{2} =
	\begin{bmatrix}
		\phi_{V+}e^{-k_xd} & \phi_{V-}e^{k_xd} \\
		D_{V+}e^{-k_xd} & D_{V-}e^{k_xd} 
	\end{bmatrix} \ ,
\end{equation*}
as we recall that both $D_{V_\pm}$ and $\phi_{V_\pm}$ are simply set by the vacuum permittivity $\epsilon_0$: $D_{V_\pm}=\pm i\epsilon_0\phi_{V_\pm}$, $\phi_{V_\pm}=1/\sqrt{\pm2i\epsilon_0}$. 

With the above definitions, the boundary conditions in Eqs.\eqref{eq:boundary_condition} can then be written in the following compact form (the detailed derivation can be found in Appendix \ref{apd:matrix_approach})
\begin{equation*}
	\pmb{M}
	\big[
	b^{(1)}_1,...b^{(1)}_4,\tilde{b}^{(2)}_1,...,\tilde{b}^{(2)}_4
	\big]^T
	=-\pmb{M}_1\pmb{U}^{(1)}_{in} b^{(1)}_{in}
\end{equation*}
where $\pmb{M}$ is a $8\times8$ matrix constructed by joining four ($\alpha=1..4$) $\pmb{M}_1\pmb{U}^{(1)}_\alpha$ and four $-\pmb{M}_2\pmb{U}^{(2)}_\alpha$ $8\times1$ column vectors together as:
\begin{equation}
	\begin{aligned}	
	\pmb{M} = \big[
	&\pmb{M}_1\pmb{U}^{(1)}_1, \pmb{M}_1\pmb{U}^{(1)}_2, \pmb{M}_1\pmb{U}^{(1)}_3, \pmb{M}_1\pmb{U}^{(1)}_4, \\ 
	&-\pmb{M}_2\pmb{U}^{(2)}_1, -\pmb{M}_2\pmb{U}^{(2)}_2, -\pmb{M}_2\pmb{U}^{(2)}_3, -\pmb{M}_2\pmb{U}^{(2)}_4
	\big] \ .
\end{aligned}
\label{eq:M}
\end{equation}


  All the reflection coefficients $r^{(1)}_\alpha\equiv b^{(1)}_\alpha/b^{(1)}_{in}$ in medium 1 and all the transmission coefficients $t^{(2)}_\alpha\equiv \tilde{b}^{(2)}_\alpha/b^{(1)}_{in}$ in medium 2 of the partial wave amplitudes can therefore be solved simultaneously as
\begin{equation}\label{eq:coef_matrix}
	\big[
	r^{(1)}_1,...r^{(1)}_4,t^{(2)}_1,...,t^{(2)}_4
	\big]^T
	= -\pmb{M}^{-1}\pmb{M}_1\pmb{U}^{(1)}_{in}.
\end{equation}

We remark that with the given materials, crystallographic orientations, the size of the vacuum gap and the frequency, the matrices $\pmb{M}$ and $\pmb{M}_1$ depend only on the wave vector component $k_x$ along the interface, which is a {\em conserved quantity} for all partial wave modes in a scattering problem. The detailed information about the incident wave, such as the normal component of the wave vector $k_z$ and the polarization $\pmb{A}_{in}$, are defined separately in the column vector $\pmb{U}^{(1)}_{in}$.  This means that the choice of the incident wave mode does not influence the calculation $\pmb{M}$ and $\pmb{M}_1$, therefore those matrices are only computed once for a given $k_x$, and the reflection and transmission coefficients for all the incident modes can readily be obtained simply by changing $\pmb{U}^{(1)}_{in}$.

In contrast, in the approach described by Ref.\cite{ALSHITS1993} in which Cramer's rule is used, the matrices used in equations (38) and (39) of Ref.\cite{ALSHITS1993} have to be re-constructed and calculated each time a new incident wave mode is given. This is because the common columns of these matrices should be the eigenvector solutions of all the transmitted wave modes except the incident mode, to ensure the columns of the fully constructed matrices are linearly independent. Furthermore, the Cramer's rule used in Ref.\cite{ALSHITS1993} requires computation of $n+1$ determinants to solve $n$ linear equations, which is considered computationally inefficient compared to the single matrix inversion used in Eq.\eqref{eq:coef_matrix} in our approach.

\subsection{Multiple reflection approach}\label{sec:multiple_reflection}
In the second, alternative approach, the reflected and transmitted waves in both media can be considered to be coupled by a superposition of multiply reflected evanescent electric potential waves in the vacuum gap. 
A similar picture has been adopted before in the description of the analogous "photon tunneling", in other words the frustrated total internal reflection phenomenon for electromagnetic waves in optics \cite{Court64,BornWolf}. 

In this approach, we define two $5\times2$ scattering matrices $\pmb{S}^{(1)}$ and $\pmb{S}^{(2)}$  for the two vacuum-solid interfaces (connecting the incoming and outgoing partial wave amplitudes for all modes), calculated separately for each interface (for the definitions, see Appendix \ref{apd:sct_matrix}, Fig.\ref{fig:smatrix} and Eq.\eqref{eq:smtx_form}). These scattering matrices are generalized in the sense that they include evanescent modes, in particular the two evanescent vacuum gap modes.   
As shown in Appendix \ref{apd:sct_matrix}, the resulting scattering matrices are
\begin{equation}\label{eq:sct_matrix}
	\begin{aligned}
		\pmb{S}^{(1)} & =  
		\begin{bmatrix}
			\bar{\pmb{r}}^{(1)} & \bar{\pmb{t}}^{(1)}\\
			\bar{t}_{{V}}^{(1)} & \bar{r}_{{V}}^{(1)}
		\end{bmatrix} \\
		& =  
		\big[\pmb{U}^{(1)}_{1},...,\pmb{U}^{(1)}_{4},-\pmb{U}_{{V_+}}\big]^{-1}
			\big[-\pmb{U}^{(1)}_{in},\pmb{U}_{{V_-}}\big]
		,\\
		\pmb{S}^{(2)} & = 
		\begin{bmatrix}
		\bar{\pmb{r}}^{(2)} & \bar{\pmb{t}}^{(2)}\\
			\bar{t}_{{V}}^{(2)} & \bar{r}_{{V}}^{(2)}
		\end{bmatrix} \\
		& =
		\big[\pmb{U}^{(2)}_{1},...,\pmb{U}^{(2)}_{4},-\pmb{U}_{{V_-}}\big]^{-1}
			\big[-\pmb{U}^{(2)}_{in},\pmb{U}_{{V_+}}\big],
	\end{aligned}
\end{equation}
where $\bar{\pmb{r}}^{(i)}=[\bar{r}^{(i)}_1 ... \bar{r}^{(i)}_4]^T$ are the reflection amplitude coefficients into modes $\alpha=1,...,4$ of the incoming wave mode from medium $i=1,2$, $\bar{r}_{{V}}^{(i)}$ the reflection amplitude coefficient from medium $i$ of an incoming vacuum mode, $\bar{\pmb{t}}^{(i)}=[\bar{t}^{(i)}_1 ... \bar{t}^{(i)}_4]^T$ the transmission amplitude coefficients of an incoming vacuum mode into wave modes $\alpha=1,...,4$ of medium $i$, and finally,  $\bar{t}_{{V}}^{(i)}$ the transmission amplitude coefficient of the incoming wave mode from medium $i$ into a vacuum mode. 
  To avoid confusion with the coupled scattering coefficients calculated with the direct approach in section \ref{sec:comb_boundaries} (Eq.\eqref{eq:coef_matrix}), we have used bars on the top of the symbols here. These "bare" coefficients describe the scattering of the electroacoustic wave as if there is no second bulk medium, and will be used below to construct the total tunneling transmission and reflection coefficients with the help of a multiple reflection factor generated by the vacuum gap. 

The total transmission factor consists of sum of partial evanescent waves in the gap that have traversed the gap once, reflected at both interfaces and traversed the gap three times, and so on. We therefore get a geometric series for the total transmission coefficient into mode $\alpha$, $t^{(2)}_{\alpha}$, as
\begin{equation*}
	\begin{aligned}
		t^{(2)}_{\alpha} =&
		\bar{t}_{V}^{(1)}e^{-k_xd} \bar{t}_{\alpha}^{(2)}
		+\bar{t}_{V}^{(1)}\bar{r}_{V}^{(1)}\bar{r}_{V}^{(2)}e^{-3k_xd}\bar{t}_{\alpha}^{(2)}\\
		&+\bar{t}_{V}^{(1)}(\bar{r}_{V}^{(1)})^2(\bar{r}_{V}^{(2)})^2e^{-5k_xd}\bar{t}_{\alpha}^{(2)}
		+\dots \\
		=&\dfrac{\bar{t}_{V}^{(1)}\bar{t}_{\alpha}^{(2)}e^{-k_xd}}{1-\bar{r}_{V}^{(1)}\bar{r}_{V}^{(2)}e^{-2k_xd}},
	\end{aligned}
\end{equation*}
where an attenuation factor $e^{-k_xd}$ due to the wave path has been included each time wave passes through the gap. We can also calculate the total reflection coefficient the same way. Collecting in both cases the common multiple reflection factor 
\begin{equation}\label{eq:multi_reflection}
f_m(d) = \dfrac{e^{-k_xd}}{1-\bar{r}_{V}^{(1)}\bar{r}_{V}^{(2)}e^{-2k_xd}},
\end{equation}
we arrive at the expressions for the total transmission and reflection coefficients $t^{(2)}_{\alpha}$ and $r^{(1)}_{\alpha}$ from the input mode into the mode $\alpha=1,...,4$ in medium (2) (transmission) or in (1) (reflection):
\begin{align}
	t^{(2)}_{\alpha} &=\bar{t}_{\alpha}^{(2)}\bar{t}_{V}^{(1)}f_m(d) \label{eq:pr_t_tot}\\
	r^{(1)}_{\alpha} &= \bar{r}_{\alpha}^{(1)}+\bar{r}_{V}^{(2)}\bar{t}_{\alpha}^{(1)}\bar{t}_{V}^{(1)}f_m(d)e^{-k_xd}. \label{eq:pr_r_tot}
\end{align}

We have checked that the results of the multiple reflection approach are identical to our previously derived scattering coefficients determined using the first, combined boundary conditions approach. The main difference is that in this second approach, the gap distance $d$ is completely separated from the calculation of the matrices. For given materials, crystal orientations and the incident wave, the scattering matrices $\pmb{S}^{(1)}$ and $\pmb{S}^{(2)}$ are independent of the gap distance, and the effects of the gap to the scattering coefficients can easily be obtained through the explicit factor $f_m(d)$. This makes the computation as a function of the gap distance easier, as the scattering matrices are computed only once. In addition, the multiple reflection approach provides an alternative physical picture of the phenomenon of tunneling of acoustic waves through a vacuum gap, analogous to the near-field electromagnetic wave "tunneling" (frustrated total internal reflection) \cite{Court64,BornWolf}. To the authors' knowledge, this multiple reflection picture  has not been described in the literature before for the problem of bulk electroacoustic wave tunneling.


\section{Illustrative examples}\label{sec:example}
\begin{figure*}[t]
	\centering
	\includegraphics[width=1\linewidth]{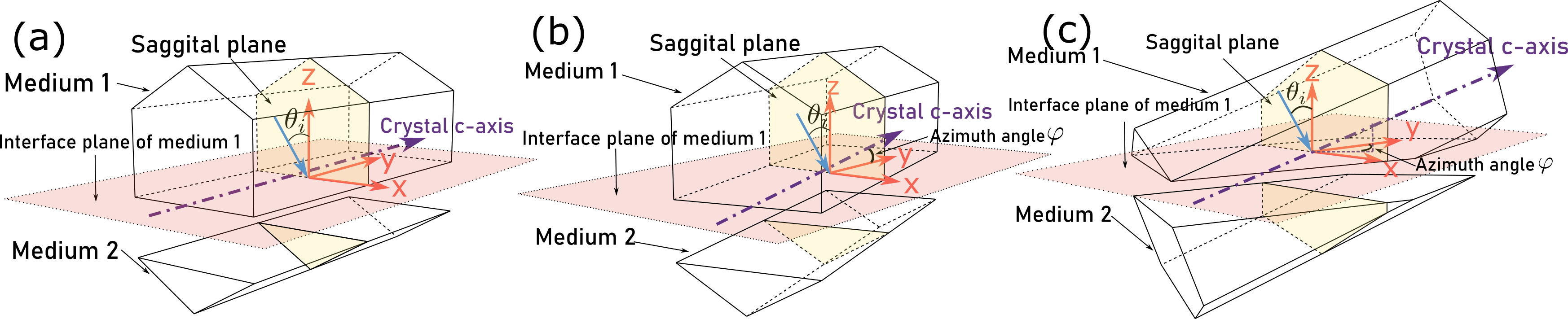}
	\caption{Wurtzite hexagonal crystal orientations used in the illustrative examples. (a) In the analytical example (section \ref{sec:analytical}), the vacuum gap is cut through the a-plane [$11\bar{2}0$] of the crystal. The azimuthal angle $\varphi=0$ and the polarization of the wave is aligned with the crystallographic $c$-axis. (b) In the first numerical example, vacuum gap is also cut through the a-plane [$11\bar{2}0$] of the crystal, while the azimuthal angle is rotated from $0^\circ$ to $360^\circ$ with respect to the $z$-axis. (c) In the second numerical example, the vacuum gap is cut through n-plane [11$\bar{2}$3] of the crystal. Crystallographic $c$-axis is no longer parallel to the interface plane, while the azimuthal angle is rotated from $0^\circ$ to $360^\circ$ with respect to the $z$-axis.}
	\label{fig:crystal_cut}
\end{figure*}
In this section, we provide some example calculations, first for a rare case that is analytically soluble. After that, we provide a  limited set of examples of numerical results for a hexagonal ZnO crystal with varying crystal orientation. The results are not meant to be exhaustive, as the main focus of this work is the introduction of the formalism and the workflow how solutions can be obtained.

\subsection{Analytical example for an incident FT bulk wave} \label{sec:analytical}
Generally speaking, for crystals with arbitrary anisotropy and orientation, it is not possible to obtain simple analytical expressions for the reflection and transmission coefficients for bulk acoustic wave tunneling. However,  for some particular incident modes and high symmetry crystal configurations, analytical solutions can be acquired. In this section, we demonstrate results for such an example: a fast transverse (FT) incident bulk wave scattering from a gap structure between two identical wurtzite hexagonal crystals (6mm symmetry) with the same crystal orientation. The acoustic polarization direction of the incident wave is aligned with the crystallographic $c-$axis, which is perpendicular to the sagittal plane (see Fig.\ref{fig:crystal_cut}(a)), in other words the c-axis is aligned with the solid-vacuum interface planes. With such a high symmetry configuration, one could also call the incident wave mode a horizontally polarized share wave (SH).   

With the above configuration, there is no mode conversion and only acoustic waves with the same polarization can be excited and scattered \cite{Auld1973}, therefore the matrix $\pmb{N}$ in the eigen-equation Eq.\eqref{eq:stroh_eigenfunc} simplifies to a $4\times4$ matrix and the eigenvectors are four-vectors $[u_y,\phi,L_{yz},D^n]^T$ (for details, see Appendix \ref{apd:analyticaldetails}):
\begin{equation}\label{eq:N4}
	\pmb{N}(v_x)=
	\begin{bmatrix}
		0 & 0 & -\frac{\epsilon_{xx}}{\epsilon_{xx} c_{44}+e_{x5}^2} & \frac{e_{x5}}{\epsilon_{xx} c_{44}+e_{x5}^2} \\
		0 & 0 & \frac{e_{x5}}{\epsilon_x c_{44}+e_{x5}^2} & \frac{c_{44}}{\epsilon_{xx} c_{44}+e_{x5}^2} \\
		c_{44}-\rho v_x^2 & -e_{x5} & 0 & 0 \\
		-e_{x5} & -\epsilon_0 & 0 & 0
	\end{bmatrix}.
\end{equation}
The phase velocity along the interface $v_x$ contained in $\pmb{N}$ can easily be found using the dispersion relation $(c_{44}+e_{15}^2/\epsilon_{11})k^2 = \rho \omega^2$ \cite{Auld1973} and the definition of the incident angle $\theta_i$ in $v_x = v \sin\theta_i$:
\begin{equation*}
	v_x^2 = \frac{\epsilon_{xx}c_{44}+e_{x5}^2}{\epsilon_{xx}\rho\sin^2\theta_i}.
\end{equation*}

A set of four eigenvalues ($p_\alpha=\pm\cot\theta_i$ and $\pm i$) and eigenvectors can be obtained, corresponding to two homogeneous (transverse modes) and two inhomogeneous (evanescent modes) partial waves. In particular, the particle displacement fields vanish in the solutions of inhomogeneous waves, but their stress fields exist. In contrast, for the bulk modes, the electrical displacement fields vanish, but not the electrical potential (Appendix \ref{apd:analyticaldetails}).

Matrix $\pmb{M}$ can be constructed following our first approach and obtained using straightforward algebra as
\begin{equation*}
	\begin{aligned}
	\pmb{M}&=\frac{1}{2}\times \\
	&\begin{bmatrix}
		-U & i(\epsilon_0-\epsilon_{xx})V & -iUe^{k_xd} & -(\epsilon_0+\epsilon_{xx})Ve^{k_xd} \\
		iU & (\epsilon_0+\epsilon_{xx})V & -Ue^{-k_xd} & i(\epsilon_0-\epsilon_{xx})Ve^{-k_xd} \\
		-i\sqrt{2\epsilon_{xx}}B & -i\sqrt{2i\epsilon_{xx}}A & 0 & 0 \\
		0 & 0 & -\sqrt{2\epsilon_{xx}}B & -\sqrt{2i\epsilon_{xx}}A
	\end{bmatrix},
	\end{aligned}
\end{equation*}
where $A=e_{x5}/\epsilon_{xx}$, $B^2=(A^2+c_{44}/\epsilon_{xx})\cot\theta_i,\{B\in\Re|B\ge0\}$, $U=i\sqrt{i\epsilon_0}A/\sqrt{\epsilon_{xx}}B$ and $V=1/\sqrt{\epsilon_0\epsilon_{xx}}$.


The exact solutions of the reflection and transmission coefficients of the fast transverse partial wave mode can be obtained from Eq.\eqref{eq:coef_matrix}:
{
\begin{align}
	r_{FT} &= 2iA^2\epsilon_0\frac{Q_+e^{2k_xd}-Q_-}
	{Q_+^2e^{2k_xd}-Q_-^2}-i \label{eq:r_SH}	\\
	t_{FT} &= -\frac{4iA^2B^2\epsilon_0\epsilon_{xx}e^{k_xd}}
	{Q_+^2e^{2k_xd}-Q_-^2}, \label{eq:t_SH}
\end{align}
}
where
\begin{equation*}
	Q_\pm=A^2\epsilon_0-iB^2(\epsilon_0\pm\epsilon_{xx}).
\end{equation*}


The alternative approach we presented in Eqs.\eqref{eq:sct_matrix}-\eqref{eq:pr_r_tot} provides identical solutions,  where the half-space scattering coefficients read as
\begin{equation*}
	\begin{aligned}
		\bar{r}_{FT}^{(1)} &=\frac{iA^2\epsilon_0-B^2(\epsilon_0+\epsilon_{xx})}{Q_+}, \\
		\bar{r}_{V}^{(1)} &= \frac{iQ_-}{Q_+}, \quad \bar{r}_{V}^{(2)} = \frac{-iQ_-}{Q_+}, \\
		\bar{t}_{V}^{(1)} &= \bar{t}_{FT}^{(1)} = \bar{t}_{FT}^{(2)} = \frac{2iAB\sqrt{i\epsilon_0\epsilon_{xx}}}{Q_+}, 
	\end{aligned}
\end{equation*}
and the multiple reflection factor is
\begin{equation}\label{eq:SH_rm}
	f_m(d)=\frac{Q_+^2e^{k_xd}}{Q_+^2e^{2k_xd}-Q_-^2}.
\end{equation}


\begin{figure*}[ht]
	\centering
	\includegraphics[width=0.7\linewidth]{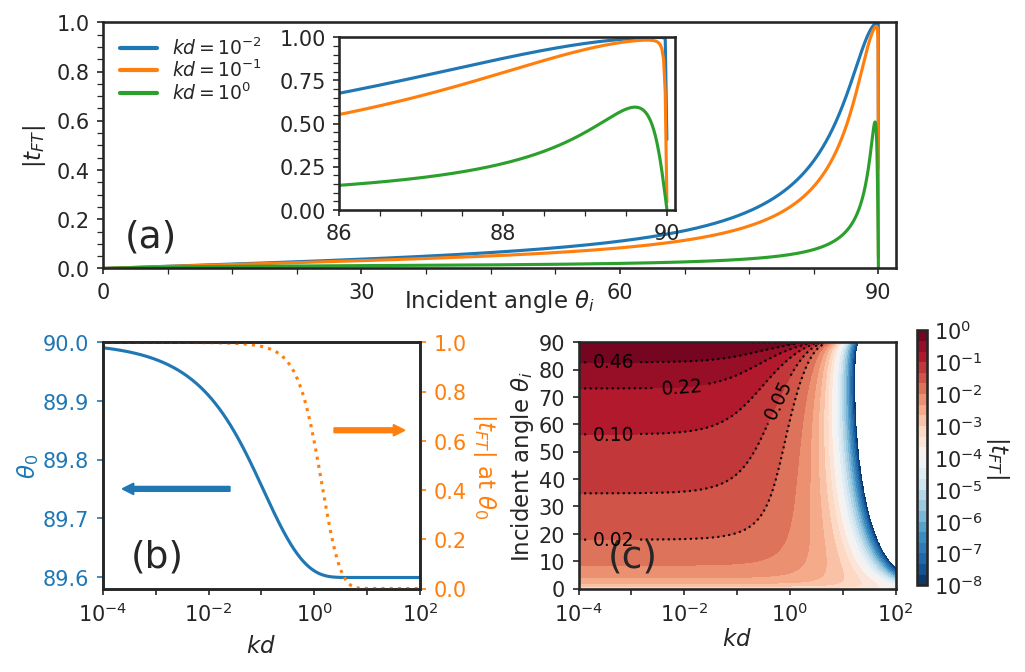}
	\caption{(a) Dependence of the magnitude of the amplitude transmission coefficients $|t_{FT}|$, from Eq.\eqref{eq:t_SH}, of a fast transverse wave tunneling between two ZnO crystals aligned as in Fig. \ref{fig:crystal_cut} (a), as a function of the incident angle $\theta_i$, with three different values of $kd=0.01,0.1,1$, where $d$ is the gap width and $k$ the incident wave vector magnitude. The inset zooms into the glancing angles. (b) Dependence of the peak transmission angle $\theta_0$ (left axis) on $kd$, with the right axis showing the corresponding $|t_{FT}|$ at $\theta_0$. (c) Colored contour plot of the magnitude of the transmission coefficient $|t_{FT}|$ in a logarithmic scale versus incident angle $\theta_i$ and $kd$.}
	\label{fig:ampvskd_SH}
\end{figure*}
Fig.\ref{fig:ampvskd_SH}(a) shows plots of the magnitudes of the tunneling transmission coefficients $|t_{FT}|$ of the fast transverse SH wave (calculated from  Eq.\eqref{eq:t_SH}) across a vacuum gap structure separating ZnO crystals with $6mm$ hexagonal symmetry oriented as in Fig.\ref{fig:crystal_cut} (a), as a function of the incident angle $\theta_i$,  with three different scaled vacuum gap values $kd$, where $k$ is the magnitude of the incident wave vector. The ZnO material constants adopted in the calculation are $\rho=5680$ kgm$^{-3}$, $c_{44}=4.247\times10^{10}$ Nm$^{-2}$, $e_{x5}=-0.48$ Cm$^{-2}$, and $\epsilon_{xx}=8.55\epsilon_0$\cite{Auld1973}. The main observation is that transmission remains modest, except at small glancing angles (near $90 ^{\circ}$ incidence), where a maximum can be found at $\theta_0$. For small enough gaps with $kd < 1$, this transmission peak approaches unity. The peak transmission condition can be found by setting the real part of the denominator in Eq.\eqref{eq:t_SH} to zero, giving an equation for $\theta_0$:
\begin{equation}\label{eq:SH_phase_match}
	B^4\equiv\frac{(A^2+c_{44}/\epsilon_{xx})^2}{\tan^2\theta_0}=\frac{A^4\epsilon_0^2(e^{2k_xd}-1)}
	{(\epsilon_0+\epsilon_{xx})^2e^{2k_xd}-(\epsilon_0-\epsilon_{xx})^2}.
\end{equation}

Furthermore, when the gap size approaches zero, the transmission coefficient, Eq.\eqref{eq:t_SH}, is simplified to the expression $t_{FT} = A^2/(A^2-iB^2)$,  which approaches unity when $\theta \rightarrow 90 ^{\circ}$.
Conversely, our expressions demonstrate that the transmission is never mathematically exactly one for a finite gap size. 

Similar analytical results for an incoming SH mode for the same crystal orientations have also been demonstrated for LiIO$_3$ (hexagonal class 6 symmetry) and Bi$_{12}$GeO$_{20}$ (cubic class 23 symmetry) by Balakirev and Gorchakov\cite{Balakirev1977}, who attributed the peak transmission to phase matching of the incident and transmitted waves. They did not, however, give explicit formulas for 6mm symmetry. For the lower class 6 symmetry of their study, they showed that a range of angles can be found for complete transmission with a finite small gap size, in contrast to our findings for the 6mm symmetry case. 

Fig.\ref{fig:ampvskd_SH}(b) gives a closer look of the angle of maximum transmission $\theta_0$ and its corresponding peak transmission value. The range of angles that can satisfy the condition is tightly limited to lie within a range of $0.4$ degrees, and with an increasing gap size beyond the characteristic length $kd\sim 1$, the peak transmission quickly drops to zero. The dependence of the transmission coefficient on the gap size is more clearly presented in Fig.\ref{fig:ampvskd_SH}(c). The bulk acoustic wave tunneling is switched off at about $kd\geq 1$, whereas the transmission is saturated for gap sizes smaller than about $kd\leq 10^{-2}$. The smallest incident angles providing a transmission factor larger than 10 \% are around $\sim 60 ^{\circ}$ for small gaps.   

\subsection{Numerical results for arbitrarily oriented hexagonal ZnO crystals}
The reflection and transmission coefficients of arbitrarily oriented crystals can also be obtained numerically by following our theoretical approach. Here, we demonstrate two sets of results (two cuts) for hexagonal 6mm ZnO crystals that been cut into two pieces and separated by a vacuum gap of distance $d$. In other words, we only consider here that the two crystals have the same orientation.  

Due to the uniaxial symmetry, the orientation of the crystals can be fixed by just two angles: the crystal zenith angle $\vartheta$ and the crystal azimuth angle $\varphi$ (see Fig.\ref{fig:Eulerangles}(b)). The details of the definition of the crystal orientation, the rotation procedure and the transformations of the material tensors are described in Appendix \ref{apd:crystal_rotation}. Both semi-infinite bulk crystals share this identical crystal orientation in which the zenith angle $\vartheta$ is fully determined by the plane of the cut (see Appendix \ref{apd:crystal_cut} for the common cut planes for a hexagonal crystal), whereas the rotation of the crystal azimuth angle $\varphi$ is equivalent to the rotational degree of freedom of the incident wave (the orientation of the sagittal plane) around the normal of the interface plane. The incident wave has two degrees of freedom: One is the incident angle $\theta_i$ that resides inside the sagittal plane and varies from $0^\circ$ to $90^\circ$. The other is the incident azimuth angle that varies from $0^\circ$ to $360^\circ$. For the cases demonstrated in this section, the rotation of the crystal azimuth angle $\varphi$ can be considered either as a change of the incident wave azimuth angle, or a change of the crystal orientation. In the computations here, we implemented it as a rotation of the crystal, to avoid duplication. 

The numerical algorithms were implemented by using the Anaconda Python distribution. Here, we briefly explain the workflow of the implementation of the combined boundary condition approach presented in section \ref{sec:comb_boundaries}. A set of input parameters specifying the material constants  (tensors $\pmb{\epsilon}^{S}_0$, $\pmb{e}_0$, $\pmb{c}^{E}_0$, and scalar $\rho$), the crystal orientation ($\vartheta$,$\varphi$), the gap distance ($d$), and the incident angle ($\theta_i$) and the mode of the incident bulk wave are first given. With the material constants and crystal orientation, the rotated material parameters ($\pmb{\epsilon}^{S}$, $\pmb{e}$, $\pmb{c}^{E}$) can be obtained by using the formulation described in Appendix \ref{apd:crystal_rotation}. 

Knowing the rotated material parameters and the incident angle and mode, the parallel component of the incident wave phase velocity $v_x$ is solved from the piezoelectrically stiffened Christoffel equations (for details, see standard textbooks, e.g. \cite{Auld1973,RoyerI}). By combining the rotated material constants and the phase velocity $v_x$, the $8\times8$ Stroh matrix $\pmb{N}(v_x)$ given by Eq.\eqref{eq:stroh_N_mtx} can be constructed, whose eigenvalues $p_\alpha$ and eigenvectors $\pmb{\xi}_\alpha$ are then solved from Eq.\eqref{eq:stroh_eigenfunc}.  The orthogonality of the eigenvectors are then checked and they are normalized using the Stroh-normalization condition, Eq.\eqref{eq:stroh_inner_basis}.

\begin{figure*}[p!]
	\centering
	\includegraphics[width=0.9\linewidth]{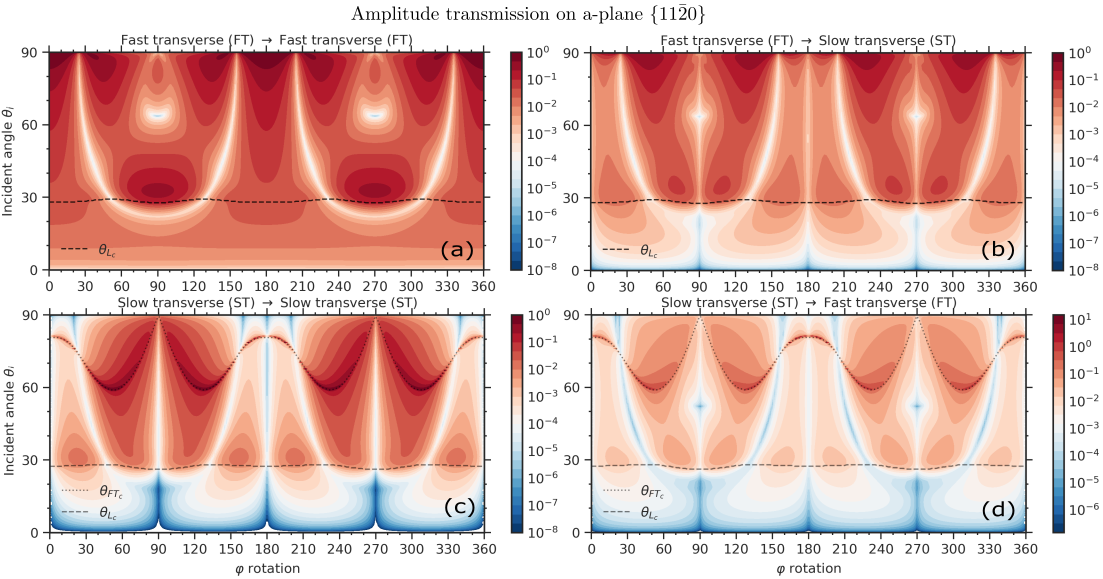}
	\caption{The magnitudes of amplitude transmission coefficients $|t_\alpha|$ across a vacuum gap (color scale) of an a-plane cut ZnO crystal ($11\overline{2}0$), versus incident angle $\theta_i$ and $z-$axis rotation angle $\varphi$, for a scaled gap $kd=10^{-2}$. Two different incident wave modes (FT, ST) and two transmitted wave modes (FT, ST) are demonstrated: (a) FT-to-FT, (b) FT-to-ST, (c) ST-to-ST, and (d) ST-to-FT transmission.  $\theta_{L_c}$ (dashed) and $\theta_{FT_c}$ (dotted) are the critical angles for scattered L and FT wave modes. Note that (d) has a different logarithmic scale, as in the  mode conversion $|t_\alpha| > 1$ is possible.}
	\label{fig:acut_transmission}
\end{figure*}

\begin{figure*}[p!]
	\centering
	\includegraphics[width=0.9\linewidth]{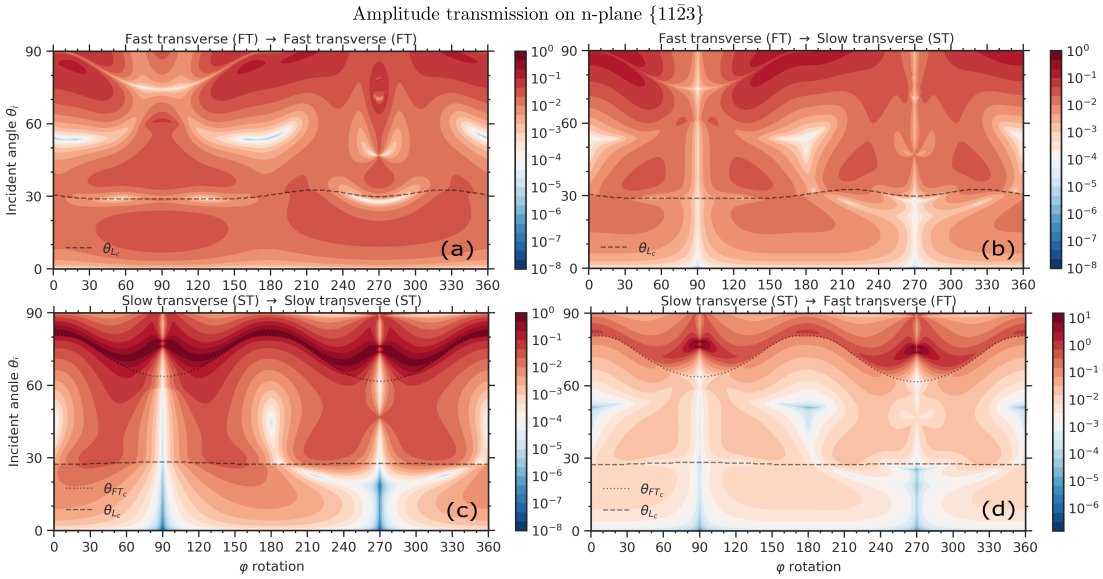}
	\caption{The magnitudes of amplitude transmission coefficients $|t_\alpha|$ across a vacuum gap (color scale) of an n-plane cut ZnO crystal ($11\overline{2}3$), versus incident angle $\theta_i$ and $z-$axis rotation angle $\varphi$, for a scaled gap $kd=10^{-2}$. Two different incident wave modes (FT, ST) and two transmitted wave modes (FT, ST) are demonstrated: (a) FT-to-FT, (b) FT-to-ST, (c) ST-to-ST, and (d) ST-to-FT transmission.  $\theta_{L_c}$ (dashed) and $\theta_{FT_c}$ (dotted) are the critical angles for scattered L and FT wave modes. Note that (d) has a different logarithmic scale, as in the  mode conversion $|t_\alpha| > 1$ is possible.}
	\label{fig:ncut_transmission}
\end{figure*}

 At this point, we can begin to solve the boundary condition problem of Eq.\eqref{eq:boundary_condition}. By following our first approach, two $8\times5$ vacuum matrices $\pmb{M}_1$ and $\pmb{M}_2$ can be constructed from Eq.\eqref{eq:vac_M} based on  $d$ and $v_x$; the $8\times8$ matrix $\pmb{M}$ can be formed by combining the vacuum matrices and the eigenvector solutions of the Stroh matrix; and the column vector $\pmb{U}^{(1)}_{in}$ can be constructed from the incident mode eigenvector $\pmb{\xi}_{in}$. Finally, with these computed matrices, the transmission and reflection coefficients of the incident bulk electroacoustic wave can be acquired from Eq.\eqref{eq:coef_matrix}. The computational time of each of the above processes took less than $1$ ms, with an overall time less than $5$ ms using a standard modern laptop. A set of results with a two varying incident angles, as shown in Figs \ref{fig:acut_transmission} and \ref{fig:ncut_transmission} then took 6 minutes each.

In the first set of results, shown in Fig.\ref{fig:acut_transmission}, we start from an $a$-plane cut crystal with $\varphi=0^\circ$, which describes an orientation that is identical to the analytical example in Section \ref{sec:analytical}. Then, we gradually rotate the crystal orientation around the $z-$axis, the normal of the interfaces, from $\varphi=0^\circ\rightarrow360^\circ$ (see Fig.\ref{fig:crystal_cut}(b)). 

We have chosen to plot just the most interesting example cases in Fig.\ref{fig:acut_transmission}, as our goal here is to demonstrate the capabilities of the formalism. We plot the magnitudes of tunneling amplitude transmission coefficients $|t_\alpha|$ of incident fast transverse (FT) and slow transverse (ST) wave modes, and their mode converted transmission amplitudes  (i.e. FT$\rightarrow$ FT, FT$\rightarrow$ ST, ST$\rightarrow$ ST ST$\rightarrow$ FT), as a function of the incident angle $\theta_i$ and the rotation angle $\varphi$, keeping the scaled gap $kd=10^{-2}$ constant. (The mode assignment process is discussed in Appendix \ref{apd:wavemode_assignment}). In comparison to the FT and ST modes, the transmission of the L mode is much weaker, does not show as many interesting features, and we choose not use it as an example here. In addition, we plot the critical incident angles, beyond which a faster reflected partial wave mode becomes evanescent.
Thus for the incident FT mode, only one critical angle exists, where the L-mode becomes evansecent ($\theta_{L_c}$), whereas for the incident ST mode, there are two critical angles: for the L-mode $\theta_{L_c}$ and for the FT mode $\theta_{FT_c}$.    

Since for this first crystal orientation example the azimuthal rotation axis $\varphi$ is perpendicular to the crystal uniaxial $c-$axis, we expect and observe a mirrored twofold symmetry in the plots. With an incident FT mode, several isolated high transmission areas are observed, and they are primarily located at small glancing angles (large $\theta_i$) around high symmetry orientations. In particular, the line segment of $\varphi=0^\circ$ and $\theta_i\in[0^\circ,90^\circ]$ for FT$\rightarrow$FT represents the same results as already discussed in the analytical calculation in Sect. \ref{sec:analytical}. However, with crystal orientations around $\varphi = 90^\circ$ and $270^\circ$, the high transmission region lies just after the critical angle $\theta_{L_c}$.  Another general observation is that for both modes, the transmission is significantly enhanced when $\theta_i$ is beyond $\theta_{L_c}$, as more energy is then concentrated near the interfaces.

With an incident ST mode, in contrast, a narrow high transmission "resonance" exists close to the critical angle of the FT partial waves ($\theta_{FT_c}$) in the first intersonic interval, and is significantly enhanced around $\varphi=n\pi/3\,,n=1,2,...$. Such a resonant transmission could be interpreted as arising from the excitation of leaky surface wave modes coupling across the gap \cite{Darinskii2006}.    
In addition to resonant features, "antiresonances", or sharp dips,  can also be observed in the transmission. In particular the u-shaped feature between $\varphi \approx 20^\circ ... 160^\circ $ and $\varphi \approx 200^\circ ... 340^\circ $ is prominent in all plots.

In the second crystal cut example of Fig.\ref{fig:ncut_transmission}, we demonstrate the same FT and ST mode results for ZnO, but which  is now initially cut from a crystallographic plane of $\{11\bar{2}3\}$ ($n$-plane, see Fig.\ref{fig:crystal_cut}(c) and Appendix \ref{apd:crystal_cut} for common cut planes for a hexagonal crystal). The change in the crystal orientation dramatically distorts the amplitude transmission as a function of both $\theta_i$ and $\varphi$. The two-fold symmetry with respect to $\varphi$ rotations is lost, and with an incident FT mode, the transmission is generally attenuated compared to the $a$-plane results. To understand this, we consider for example the case $\varphi=0$ in the $n$-plane crystal cut, for which the incident FT mode is a quasi-transverse mode which now couples to all other acoustic modes at the interface. As a result, the FT$\rightarrow$FT transmission is attenuated, while the mode converted FT$\rightarrow$ST transmission increases. In contrast, with the $a$-plane cut the incident FT mode wave is a pure horizontal shear wave (SH), as described in the analytical example, leading intuitively to a stronger FT$\rightarrow$FT transmission and vanishing FT$\rightarrow$ST transmission. Furthermore, it is interesting to see that with an incident ST mode, significant transmission resonance just beyond the FT wave critical angle $\theta_{FT_c}$ still survives as a robust feature also for the n-cut. As mentioned above, this can be interpreted as excitation of coupled leaky surface waves between the vacuum interfaces\cite{Darinskii2006}.  

\section{Conclusions and outlook}
\label{sec:conclusions}

We have shown that in general, bulk acoustic waves can be transmitted ("tunnel") across a finite vacuum gap  between two piezoelectric crystals. This mechanism works not only in the nanoscale, but also for large gap widths of the order of the wavelength. Although the effect is known in literature for some particular cases, no rigorous general formulation to study it has been put forward before. Here, we presented an approach and formalism that can be applied to study this effect for any anisotropic piezoelectric crystals with arbitrary crystallographic orientation, acquiring the solutions of reflection and transmission coefficients of all the partial waves. The extended Stroh formalism, briefly reviewed for the benefit of the reader, was used as a powerful tool to solve in general the scattering of an electroacoustic wave on the solid-vacuum interface. Two new approaches to solve the reflection and transmission coefficients of the coupled tunneling problem (two interfaces separated by a gap) were then derived: one based on the direct solution of the boundary conditions, the other on the physical picture of multiple reflections of evanescent waves in the vacuum gap. In particular, the multiple reflection method provides a physical insight of the acoustic tunneling that is analogous to near-field tunneling of evanescent electromagnetic waves. In this picture, the effect of the vacuum gap size on the reflection and transmission coefficients is conveniently separated and described by a single multiple reflection factor, offering a potential computational advantage. 

To verify the usefulness and validity of the methodology, explicit example solutions for the case of two adjacent ZnO wurtzite hexagonal crystals were demonstrated. First, we presented analytical results for a fast transverse incident mode and high symmetry crystal orientation.  Simple expressions for the transmission and reflection coefficients and the multiple reflection factor were derived, and an explicit mathematical condition for the peak transmission was also presented. We made the observation that tunneling transmission is not necessarily small: For small glancing angle incidence, transmission was approaching one for gap sizes smaller than the wavelength. 

Second, we described the workflow for numerical implementation for an arbitrary orientation, and presented some numerical results for two cases of anisotropic ZnO crystals (two different crystal cut surfaces). We plotted the transmission coefficients of the fast and slow transverse partial modes, as well as the conversion between them, against the incident angle and the crystal azimuth rotation angle. In the numerical examples, we also find close to unity transmissions, and not only for small glancing angles. Such cases were mostly observed in the vicinity of the critical angles of the scattered partial wave modes, where they become inhomogeneous surface modes. The enhancement of tunneling transmission was a particularly sharp and strong feature (resonance) for an incoming slow transverse wave with an incident angle just beyond the critical angle for the fast transverse wave, where coupled leaky surface waves can be excited.  

With the formalism and the approaches derived in this work, we have set the foundation for  many further studies of electroacoustic wave tunneling. The first straightforward objective is to map and understand the conditions for  exceptionally high transmission, as there are indications of the possibility of complete acoustic wave tunneling. In addition to direct applications in the manipulation of acoustic waves, our formalism can be applied in the future in other areas of physics related to vibrations, such as heat transport, optomechanics and quantum information science.    

\begin{acknowledgments}
	This study was supported by the Academy of Finland project number 341823. We wish to acknowledge discussions with Dr. Tuomas Puurtinen.
\end{acknowledgments}

\appendix
\section{Plane wave equations of motion and constitutive equations in quasistatic approximation}\label{apd:voigtnotation}
Here we clarify the definitions of the variables in Eqs.\eqref{eq:const_equs}, presented in the abbreviated (Voigt) matrix notation. 
The first equation in the set, the acoustic field equation, reads in general (if no external body forces are present)
\begin{equation*}
		\nabla\cdot\pmb{\sigma} = \rho\frac{\partial^2 \pmb{u}}{\partial t^2},
	\end{equation*}  
where $\pmb{\sigma}$ is the stress tensor and $\pmb{u}$ the displacement vector. Transforming it into the abbreviated Voigt notation, where the capital Voigt index $K$ runs over six coordinate pairs $K = xx,yy,zz,yz,xz,xy$, will result in a matrix equation $\nabla_{iK}\sigma_{K} = \rho\frac{\partial^2 u_i}{\partial t^2}$ where the index $i$ denotes the usual Cartesian component and repeated indices are summed. Thus, $\nabla_{iK}$ defines a $3\times6$ differential operator matrix (see for example Ref.\cite{Auld1973} Eq. (2.36) for an explicit expression). For harmonic plane waves, such an operator is replaced by a $3\times6$ matrix formed by the wave vector components $-ik_{iK}$, explicitly defined as \cite{Auld1973}
\begin{equation}
\hat{\pmb{k}} = \begin{bmatrix}
		k_x & 0 & 0 & 0 & k_z & k_y \\
		0 & k_y & 0 & k_z & 0 & k_x \\
		0 & 0 & k_z & k_y & k_x & 0 
	\end{bmatrix},
	\end{equation} 
 and the second derivative w.r.t time is replaced by $-\omega^2$, yielding the first equation in Eq.\eqref{eq:const_equs} in the main text,
\begin{equation}
	ik_{iK}\sigma_{K} = \rho\omega^2u_i.	
	\end{equation}       
 
The second equation, Gauss's law $\nabla \cdot \pmb{D} = 0$, is the only Maxwell's equation that needs to be satisfied within the quasistatic approximation. It contains the usual vector divergence operator, and it can directly be written in the component form as  $\nabla_i D_i = 0$, which gives for the plane waves the equation in the main text,
 \begin{equation}
	-ik_iD_i = 0,	
	\end{equation}    
where $k_i$ are now simply the Cartesian components of the wave vector. 

The third and the fourth equations in Eqs.\eqref{eq:const_equs} are the constitutive relations for piezoelectrics, coupling the elastic and electric variables. They are given in abbreviated notation for the stress as $\sigma_K = c_{KL}^{E}s_L-e_{Kj}E_j$ and for the electric displacement $D_i = \epsilon_{ij}^SE_j + e_{iL}s_L$ , where $s_L$ is the strain tensor, $E_i$ the electric field, and $c_{KL}^{E}$, $e_{Kj}$ and $\epsilon_{ij}^S$ are the material parameters: the elastic stiffness tensor at constant electric field, the piezoelectric strain tensor, and the electric permittivity tensor at constant strain, respectively. The constitutive relations simplify in the quasistatic case by writing them in terms of the displacement $u_i$ and the electric potential $\varPhi$, $s_L=\nabla_{Lj}u_j$, $E_i = -\nabla_i\varPhi$, and as before for the plane waves the differential operators can be substituted by  $\nabla_{Lj} \rightarrow -ik_{Lj}$, $\nabla_i \rightarrow -ik_i$, which lead to the forms presented in Eqs.\eqref{eq:const_equs}:
\begin{equation}
	\begin{aligned}
	\sigma_{K} &= -ic^E_{KL}k_{Lj}u_{j}-ie_{Kj}k_j\varPhi \\
		D_i &= -ie_{iL}k_{Lj}u_j + i\epsilon^S_{ij}k_j\varPhi. \\
	\end{aligned}
\end{equation}

\section{Extended Stroh formalism}\label{apd:stroh_formalism}
In this appendix, we provide the derivation for the piezoelectric Stroh eigen-equation, Eq.\eqref{eq:stroh_eigenfunc}, in the quasistatic approximation, including all necessary definitions, and provide a few remarks about the normalization.  
Under the framework of the quasistatic approximation, one can derive the normal projections of the stress and electric displacement fields using the piezoelectric constitutive equations in Eq.\eqref{eq:const_equs} as:
\begin{equation}
	\begin{aligned}
		\pmb{n}\cdot\pmb{\sigma} & = -in_{iK}c^E_{KL}k_{Lj}u_{j}-in_{iK}e_{Kj}k_j\varPhi \\
		\pmb{n}\cdot\pmb{D} &= -in_ie_{iL}k_{Lj}u_j + in_i\epsilon^S_{ij}k_j\varPhi \ ,\\
	\end{aligned}	
	\label{eq:stroh_constitutive_field}
\end{equation}
where $n_i$ is the i-th Cartesian component of the inward unit normal vector $\pmb{n}$ of the piezo-vacuum surface, and the $3\times6$ matrix $n_{iK}$ has the same structure as $k_{iK}$, but is now formed by the unit normal components $n_i$, explicitly written as
\begin{equation}
\hat{\pmb{n}} = \begin{bmatrix}
		n_x & 0 & 0 & 0 & n_z & n_y \\
		0 & n_y & 0 & n_z & 0 & n_x \\
		0 & 0 & n_z & n_y & n_x & 0 
	\end{bmatrix}.
	\end{equation}  
If we introduce a matrix expression $(\pmb{n}\pmb{k})$ for a $4\times4$ matrix that is defined as
\begin{equation}
	(\pmb{n}\pmb{k}) \equiv
	\begin{bmatrix}
		n_{iK} & 0 \\
		0 & n_i
	\end{bmatrix}
	\begin{bmatrix}
		c_{KL}^E & e_{Kj} \\
		e_{iL} & -\epsilon_{ij}^S
	\end{bmatrix}
	\begin{bmatrix}
		k_{Lj} & 0 \\
		0 & k_j
	\end{bmatrix} \ ,
	\label{eq:matrixdef}
\end{equation}
where the elements of the matrices represent sub-matrices instead of scalars ($n_{iK}$ represents the matrix $\hat{\pmb{n}}$, etc.), 
it is straightforward to show that Eqs.\eqref{eq:stroh_constitutive_field} can be written in the following more compact notation:
\begin{equation}
	\begin{aligned}
		\begin{bmatrix}
			\pmb{n}\cdot\pmb{\sigma} \\
			\pmb{n}\cdot\pmb{D}
		\end{bmatrix}
		=-i(\pmb{n}\pmb{k})
		\begin{bmatrix}
			\pmb{u} \\
			\varPhi
		\end{bmatrix} \ .
	\end{aligned}
	\label{eq:stroh_top_1}
\end{equation}

By adopting the general field solutions for a mode $\alpha$ in Eqs.\eqref{eq:stroh_field_functions} and decomposing the k-vector $\pmb{k} = k_x(\pmb{m}+p_\alpha\pmb{n})$ using the two orthogonal unit vectors $\pmb{m}$ and $\pmb{n}$, where $\pmb{m}$ is parallel to the piezo-vacuum interfaces,  we can further arrange Eq.\eqref{eq:stroh_top_1} into a form where the unknown $p_\alpha$ is separated into the right-hand side of the equation:
\begin{equation}
	-(\pmb{n}\pmb{n})^{-1}(\pmb{n}\pmb{m})
	\begin{bmatrix}
		\pmb{A}_\alpha \\
		\phi_\alpha
	\end{bmatrix}
	-(\pmb{n}\pmb{n})^{-1}
	\begin{bmatrix}
		\pmb{L}_\alpha \\
		D_\alpha^n
	\end{bmatrix}
	=
	p_\alpha
	\begin{bmatrix}
		\pmb{A}_\alpha \\
		\phi_\alpha
	\end{bmatrix} \ .
	\label{eq:stroh_top_2}
\end{equation}
The matrices $(\pmb{n}\pmb{n})$ and $(\pmb{n}\pmb{m})$ are defined analogously to Eq.\eqref{eq:matrixdef},  
and thus depend only on the material parameters and the orientation of the crystal. We note that real materials do not present any pathological cases where the matrix inverse $(\pmb{n}\pmb{n})^{-1}$ would not exist.  

Furthermore, the equation of motion and Gauss's law can also be organized into a similar linear equations set:
\begin{equation}
	\begin{aligned}
		(\pmb{k}\pmb{k})
		\begin{bmatrix}
			\pmb{u} \\
			\varPhi
		\end{bmatrix}
		=\rho \omega^2\pmb{I}'
		\begin{bmatrix}
			\pmb{u} \\
			\varPhi
		\end{bmatrix} \ ,
	\end{aligned}
	\label{eq:stroh_bot_1}
\end{equation}
in which 
 $\pmb{I}'$ is a $4\times4$ matrix with elements $I_{ii}^{'} = 1$, $i=1,2,3$ and others zero. 

By decomposing the k-vector as above and substituting the expression for $p_\alpha [\pmb{A}_\alpha, \varPhi_\alpha]^{T}$ from Eq.\eqref{eq:stroh_top_2}, we obtain 
\begin{widetext}
\begin{equation}
		-\left[
		(\pmb{m}\pmb{n})(\pmb{n}\pmb{n})^{-1}(\pmb{n}\pmb{m})-(\pmb{m}\pmb{m})+\rho v_x^2\hat{\pmb{I}}'
		\right]
		\begin{bmatrix}
			\pmb{A}_\alpha \\
			\phi_\alpha
		\end{bmatrix}
		-(\pmb{m}\pmb{n})(\pmb{n}\pmb{n})^{-1}
		\begin{bmatrix}
			\pmb{L}_\alpha \\
			D_\alpha^n
		\end{bmatrix} 
		=
		p_\alpha
		\begin{bmatrix}
			\pmb{L}_\alpha \\
			D_\alpha^n
		\end{bmatrix}  \ ,
	\label{eq:stroh_bot_2}
\end{equation}
\end{widetext}
where $v_x=\omega/k_x$. 
Finally, combining Eqs.\eqref{eq:stroh_top_2} and Eqs.\eqref{eq:stroh_bot_2} by defining an eight-dimensional eigenvector $\pmb{\xi}_\alpha=[\pmb{A}_\alpha,\phi_\alpha,\pmb{L}_\alpha,D_\alpha^n]^T$, we have derived the eigenequation for the piezoelectric scattering problem,  Eq.\eqref{eq:stroh_eigenfunc}:
\begin{equation}
	\pmb{N}(v_x)\pmb{\xi}_\alpha = p_\alpha\pmb{\xi}_\alpha \ ,
\end{equation}
where 
the $8\times8$ real matrix $\pmb{N}$ reads
\begin{equation}
	\begin{aligned}	
	&\pmb{N}(v_x) = \\
	&-\begin{bmatrix}
		(\pmb{n}\pmb{n})^{-1}(\pmb{n}\pmb{m}) & (\pmb{n}\pmb{n})^{-1} \\
		(\pmb{m}\pmb{n})(\pmb{n}\pmb{n})^{-1}(\pmb{n}\pmb{m})-(\pmb{m}\pmb{m})+\rho v_x^2\pmb{I}^{'} & 
		(\pmb{m}\pmb{n})(\pmb{n}\pmb{n})^{-1}
	\end{bmatrix}.
\end{aligned}
\label{eq:stroh_N_mtx}
\end{equation}

The set of eigenvectors $\xi_\alpha$ are orthogonal and form a complete set in the usual case, where the eigenvalues $p_\alpha$ are distinct \cite{Lothe1976}. In few isolated situations, non-semisimple degeneracy can occur, in which case generalized eigenvectors can be introduced \cite{Chadwick1977,Darinskii2003}. We do not consider those special cases (transonic states) in this study, as numerically one can always solve the problem in a limiting manner very close to such a special point.  

In the main text, we quoted the orthonormalization condition in Eq.\eqref{eq:stroh_inner_basis}. It follows \cite{Chadwick1977,Ting1996book} from the symmetry condition for the auxiliary matrix $\hat{\pmb{T}}\pmb{N}$ 	
\begin{equation*}
	(\hat{\pmb{T}}\pmb{N})^T=\hat{\pmb{T}}\pmb{N} \ ,
\end{equation*}
where
\begin{equation}
	\hat{\pmb{T}} =
	\begin{bmatrix}
		\hat{\pmb{O}}_{(4)} & \hat{\pmb{I}}_{(4)} \\
		\hat{\pmb{I}}_{(4)} & \hat{\pmb{O}}_{(4)} \\
	\end{bmatrix}\ ,
	\label{eq:stroh_T_mtx}
\end{equation}

with which a reciprocal eigenvector set $\pmb{T\xi}_\alpha$ orthogonal to $\pmb{\xi}_\alpha$ can be defined. It follows that the eigenvectors satisfy the relation
\begin{equation}
	\pmb{\xi}_\alpha\cdot\hat{\pmb{T}}\pmb{\xi}_\beta 
	 = \pmb{\xi}_\alpha^T\hat{\pmb{T}}\pmb{\xi}_\beta 
	= \delta_{\alpha\beta},\ \alpha,\beta=1,...,8 \ ,\label{eq:stroh_inner_basis_2}
\end{equation}
where $\delta_{\alpha\beta}$ is the Kronecker delta. We stress here that the dot symbol has the meaning of a matrix product in this context, as exemplified by the second form. In particular, it does not denote a complex inner product, in which case complex conjugation would be included for one of the vectors. 

Eq.\eqref{eq:stroh_inner_basis_2} (identical to Eq.\eqref{eq:stroh_inner_basis} in the main text) is thus readily available as the orthonormalization condition to secure an unique normalized eigenvector solution for the $\pmb{N}$ matrix. In addition, the orthonormalization condition Eq.\eqref{eq:stroh_inner_basis_2} leads to a completeness condition
\begin{equation}
	\sum_{\alpha}
	\pmb{\xi}_\alpha\otimes\hat{\pmb{T}}\pmb{\xi}_\alpha = \hat{\pmb{I}}_{(8)},
\end{equation}
providing us a powerful tool to ensure the accuracy of the solutions.

\section{Matrix solution of the boundary conditions}\label{apd:matrix_approach}

In this appendix, to avoid cumbersome expressions $D_\alpha^{n,(i)}$ and $D_{V_\pm}^n$ with the explicit superscript $n$, we will use the short-hand notation $D_\alpha^{(i)}$ and $D_{V_\pm}$ in their place to represent the Stroh eigenvector components for the normal projection of the electric displacement in Eq.(2) in the solid $i=1,2$ and in vacuum (V), respectively.

The boundary conditions of the tunneling problem, Eqs.\eqref{eq:boundary_condition}, provide a total of ten linear equations corresponding to ten partial wave amplitude solutions ($b^{(1)}_1,...,b^{(1)}_4,\tilde{b}^{(2)}_1,...,\tilde{b}^{(2)}_4,b_{V\pm}$). Here, we list explicitly all the boundary conditions included in Eqs.\eqref{eq:boundary_condition} :
\begin{equation}
	\begin{aligned}
		b^{(1)}_{in}\phi^{(1)}_{in}+\sum_{\alpha=1}^{4}b^{(1)}_\alpha\phi^{(1)}_\alpha &= b_{V+}\phi_{V+}+b_{V-}\phi_{V-} \\
		b^{(1)}_{in}D^{(1)}_{in}+\sum_{\alpha=1}^{4}b^{(1)}_\alpha D^{(1)}_\alpha &= b_{V+}D_{V+}+b_{V-}D_{V-} \\
		b^{(1)}_{in}\pmb{L}^{(1)}_{in}+\sum_{\alpha=1}^{4}b^{(1)}_\alpha\pmb{L}^{(1)}_\alpha &= 
		\hat{\pmb{O}}_{(3)} 
		\\
		\sum_{\alpha=1}^{4}\tilde{b}^{(2)}_\alpha\phi^{(2)}_\alpha &= b_{V+}\phi_{V+}e^{-k_xd}+b_{V-}\phi_{V-}e^{k_xd} \\
		\sum_{\alpha=1}^{4}\tilde{b}^{(2)}_\alpha D^{(2)}_\alpha
		&= b_{V+}D_{V+}e^{-k_xd}+b_{V-}D_{V-}e^{k_xd} \\
		\sum_{\alpha=1}^{4}\tilde{b}^{(2)}_\alpha\pmb{L}^{(2)}_\alpha &= \hat{\pmb{O}}_{(3)}. \\
	\end{aligned}
	\label{eq:boundary_condition_explicit}
\end{equation}

The goal of this appendix is to rearrange the above boundary conditions into a simple matrix equation that separates the incident wave properties, the information about the materials properties and the vacuum gap, and the scattered amplitudes, i.e. by writing 
\begin{equation}
	\hat{y} = \pmb{M}\hat{x},
	\label{eq:simple_matrix}
\end{equation}
where $\hat{y}$ is a $8\times1$ column vector that contains the information about the incident wave and $\pmb{M}$ is a $8\times8$ matrix, both to be derived below, and $\hat{x}$ and is a $8\times1$ column vector containing the wave amplitudes of all scattered waves (and therefore the information on the transmission and reflection coefficients):
\begin{equation*}
	\hat{x} = [b^{(1)}_1,...b^{(1)}_4,\tilde{b}^{(2)}_1,...,\tilde{b}^{(2)}_4]^T.
\end{equation*}

By eliminating $b_{V\pm}$ from Eq.\eqref{eq:boundary_condition_explicit}, the number of linear equations provided by the boundary conditions in Eq.\eqref{eq:boundary_condition_explicit} can be reduced from 10 to 8:

\begin{equation}
	\begin{aligned}
		\pmb{V}_1^{-1}
		\bigg(
		\begin{bmatrix}
			\phi^{(1)}_{in} \\
			D^{(1)}_{in}
		\end{bmatrix} b^{(1)}_{in}
		+\sum_{\alpha=1}^{4}
		&\begin{bmatrix}
			\phi^{(1)}_{\alpha} \\
			D^{(1)}_{\alpha}
		\end{bmatrix}b^{(1)}_\alpha
		\bigg)  \\
		& = \pmb{V}_2^{-1}
		\bigg(
		\sum_{\alpha=1}^{4}
		\begin{bmatrix}
			\phi^{(2)}_{\alpha} \\
			D^{(2)}_{\alpha}
		\end{bmatrix}\tilde{b}^{(2)}_\alpha
		\bigg) 
		\\
		b^{(1)}_{in}\pmb{L}^{(1)}_{in}+\sum_{\alpha=1}^{4}b^{(1)}_\alpha\pmb{L}^{(1)}_\alpha 
		& = \hat{\pmb{O}}_{(3)} 
		\\
		\sum_{\alpha=1}^{4}\tilde{b}^{(2)}_\alpha\pmb{L}^{(2)}_\alpha 
		& = \hat{\pmb{O}}_{(3)} \ , \\
	\end{aligned}
	\label{eq:boundary_condition_reduced}
\end{equation}

where
\begin{equation*}
	\pmb{V}_{1} =
	\begin{bmatrix}
		\phi_{V+} & \phi_{V-} \\
		D_{V+} & D_{V-} 
	\end{bmatrix} \ ,
	\pmb{V}_{2} =
	\begin{bmatrix}
		\phi_{V+}e^{-k_xd} & \phi_{V-}e^{k_xd} \\
		D_{V+}e^{-k_xd} & D_{V-}e^{k_xd} 
	\end{bmatrix} \ .
\end{equation*}
The $2\times2$ matrices $\pmb{V}_{1}$ and $\pmb{V}_{2}$ are not dependent on the incoming or scattered wave properties 
except for the conserved wave vector component $k_x$.

To combine all equations in Eqs.\eqref{eq:boundary_condition_reduced} into one matrix equation, we can move all the terms depending on $b^{(1)}_{in}$ to the right and all the others to the left, and write all equations in terms of $5\times1$ column vectors $\pmb{U}_\gamma^{(i)}=[\phi_\gamma^{(i)},D_\gamma^{(i)},\pmb{L}^{(i)}_\gamma]^T$   containing the reflected ($\pmb{U}_\alpha^{(1)}$), transmitted ($\pmb{U}_\alpha^{(2)}$) and input wave ($\pmb{U}_{in}^{(1)}$) Stroh eigenvector components for the electric potential, electric displacement and traction force. This way we obtain

\begin{widetext}
\begin{equation}
	\begin{aligned}
		\sum_{\alpha=1}^{4}\bigg(
		\begin{bmatrix}
			\pmb{V}_1^{-1} & \hat{\pmb{O}}_{(2\times3)}
		\end{bmatrix}
		\pmb{U}_\alpha^{(1)} b^{(1)}_\alpha
		-
		\begin{bmatrix}
			\pmb{V}_2^{-1} & \hat{\pmb{O}}_{(2\times3)}
		\end{bmatrix}
		\pmb{U}_\alpha^{(1)} \tilde{b}^{(2)}_\alpha 
		\bigg)
		&= 
		-\begin{bmatrix}
			\pmb{V}_1^{-1} & \hat{\pmb{O}}_{(2\times3)}
		\end{bmatrix}
		\pmb{U}_{in}^{(1)} b^{(1)}_{in}
		\\
		\sum_{\alpha=1}^{4}
		\bigg(
		\begin{bmatrix}
			\hat{\pmb{O}}_{3\times2} & \hat{\pmb{I}}_{(3)}
		\end{bmatrix}
		\pmb{U}_\alpha^{(1)} b^{(1)}_\alpha 
		-
		\begin{bmatrix}
			\hat{\pmb{O}}_{3\times2} & \hat{\pmb{O}}_{(3\times3)}
		\end{bmatrix}
		\pmb{U}_\alpha^{(2)} \tilde{b}^{(2)}_\alpha
		\bigg) 
		&= 
		-\begin{bmatrix}
			\hat{\pmb{O}}_{3\times2} & \hat{\pmb{I}}_{(3)}
		\end{bmatrix}
		\pmb{U}_{in}^{(1)} b^{(1)}_{in}
		\\
		\sum_{\alpha=1}^{4}
		\bigg(
		\begin{bmatrix}
			\hat{\pmb{O}}_{3\times2} & \hat{\pmb{O}}_{(3\times3)}
		\end{bmatrix}
		\pmb{U}_\alpha^{(1)} b^{(1)}_\alpha
		-
		\begin{bmatrix}
			\hat{\pmb{O}}_{3\times2} & \hat{\pmb{I}}_{(3)}
		\end{bmatrix}
		\pmb{U}_\alpha^{(2)} \tilde{b}^{(2)}_\alpha
		\bigg)
		&= 
		-\begin{bmatrix}
			\hat{\pmb{O}}_{3\times2} & \hat{\pmb{O}}_{(3\times3)}
		\end{bmatrix}
		\pmb{U}_{in}^{(1)} b^{(1)}_{in},
	\end{aligned}
	\label{eq:stacked_matrix_explicit}
\end{equation}
\end{widetext}
where $\hat{\pmb{O}}_{n\times m}$ denotes a zero matrix of dimensions $n\times m$, and $\hat{\pmb{I}}_{(3)}$ is the $3\times3$ identity matrix.  	

From the above form, Eqs.\eqref{eq:stacked_matrix_explicit}, we see that a single matrix equation 
\begin{equation}
	\sum_{\alpha=1}^4\big(
	\pmb{M}_1\pmb{U}^{(1)}_\alpha b^{(1)}_\alpha - \pmb{M}_2\pmb{U}^{(2)}_\alpha\tilde{b}^{(2)}_\alpha
	\big)
	=-\pmb{M}_1\pmb{U}^{(1)}_{in} b^{(1)}_{in} \
	\label{eq:stacked_matrix}
\end{equation}
can be written, if we define the $8\times5$  
matrices $\pmb{M}_1$ and $\pmb{M}_2$ as
\begin{equation}
	\pmb{M}_1 =
	\begin{bmatrix}
		\pmb{V}_1^{-1} & \hat{\pmb{O}}_{(2\times3)} \\
		\hat{\pmb{O}}_{(3\times2)} & \hat{\pmb{I}}_{(3)} \\
		\hat{\pmb{O}}_{(3\times2)} & \hat{\pmb{O}}_{(3\times3)} \\
	\end{bmatrix} \ ,
	\pmb{M}_2 =
	\begin{bmatrix}
		\pmb{V}_2^{-1} & \hat{\pmb{O}}_{(2\times3)} \\
		\hat{\pmb{O}}_{(3\times2)} & \hat{\pmb{O}}_{(3\times3)} \\
		\hat{\pmb{O}}_{(3\times2)} & \hat{\pmb{I}}_{(3)} \\
	\end{bmatrix} \ .	
\end{equation}

Finally, by comparing Eq.\eqref{eq:stacked_matrix} with the targeted expression $\hat{y} = \pmb{M}\hat{x}$ (Eq.\eqref{eq:simple_matrix}, remembering that $\hat{x} = [b^{(1)}_1,...b^{(1)}_4,\tilde{b}^{(2)}_1,...,\tilde{b}^{(2)}_4]^T)$, we obtain $\hat{y}=-\pmb{M}_1\pmb{U}_{in}b_{in}$, with the $8\times8$ matrix $\pmb{M}$ formed  by combining eight $8\times1$ column matrix blocks as

\begin{equation*}
	\begin{aligned}	
	\pmb{M} = \big[
	&\pmb{M}_1\pmb{U}^{(1)}_1, \pmb{M}_1\pmb{U}^{(1)}_2, \pmb{M}_1\pmb{U}^{(1)}_3, \pmb{M}_1\pmb{U}^{(1)}_4, \\ 
	&-\pmb{M}_2\pmb{U}^{(2)}_1, -\pmb{M}_2\pmb{U}^{(2)}_2, -\pmb{M}_2\pmb{U}^{(2)}_3, -\pmb{M}_2\pmb{U}^{(2)}_4
	\big] \ ,
\end{aligned}
\end{equation*}
which is the definition given in the main text in Eq.\eqref{eq:M}. From the above, we see that in the  matrix $\pmb{M}$, the submatrices $\pmb{M}_1$ and $\pmb{M}_2$ depend only on the vacuum permittivity $\epsilon_0$ ($D_{V_\pm}=\pm i\epsilon_0\phi_{V_\pm}$ and $\phi_{V_\pm}=1/\sqrt{\pm2i\epsilon_0}$, section \ref{sec:tunneling} the main text), the gap distance $d$ and the conserved incident wave k-vector component $k_x$; vectors $\pmb{U}_\gamma$ are the physical solutions obtained from the eigenvectors $\pmb{\xi}_\gamma$ in the extended Stroh formalism.

Finally, the four reflection and four transmission coefficients of the wave amplitudes can readily be obtained as:
\begin{equation}
	\begin{bmatrix}
		\pmb{r}_{(4\times1)} \\
		\pmb{t}_{(4\times1)}
	\end{bmatrix}
	=
	\frac{\hat{x}}{b_{in}}
	= -\pmb{M}^{-1}\pmb{M}_1\pmb{U}_{in}.
\end{equation} 

\section{Scattering matrix}\label{apd:sct_matrix}
\begin{figure}[h]
	\centering
	\includegraphics[width=0.6\linewidth]{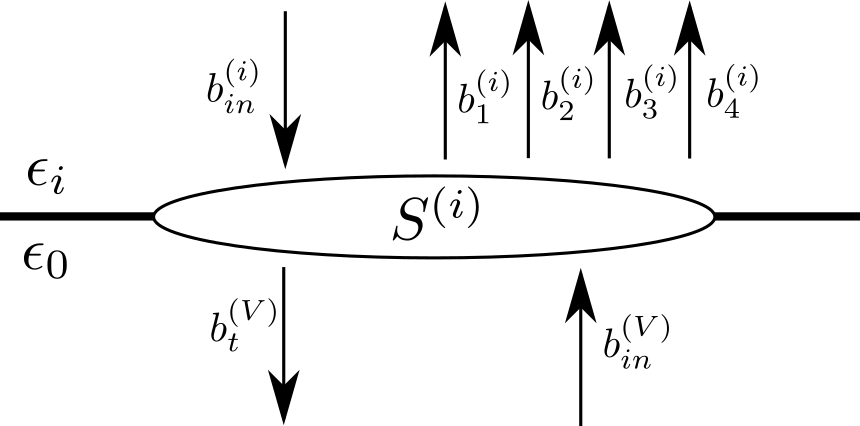}
	\caption{Illustration of the scattering matrix $\pmb{S}^{(i)}$ for an interface between medium $i$ and vacuum. For the second medium on the receiving side, the incoming amplitude $b_{in}^{(i)}$ is zero. }
	\label{fig:smatrix}
\end{figure}
If we consider a scattering problem for an interface $i$ between a piezoelectric crystal and vacuum, as illustrated in Fig.\ref{fig:smatrix}, the scattering matrix $S^{(i)}$ determining how input waves scatter into output waves can be defined as
\begin{equation}\label{eq:smtx_form}
	\begin{bmatrix}
		b_1^{(i)} \\
		b_2^{(i)} \\
		b_3^{(i)} \\
		b_4^{(i)} \\
		b_t^{(V)} \\
	\end{bmatrix}
	=S^{(i)}
	\begin{bmatrix}
		b_{in}^{(i)} \\
		b_{in}^{(V)} \\
	\end{bmatrix}
	=
	\begin{bmatrix}
		\bar{r}_1^{(i)} & \bar{t}_1^{(i)} \\
		\bar{r}_2^{(i)} & \bar{t}_2^{(i)} \\
		\bar{r}_3^{(i)} & \bar{t}_3^{(i)} \\
		\bar{r}_4^{(i)} & \bar{t}_4^{(i)} \\
		\bar{t}_{in}^{(i)} & \bar{r}_{in}^{(i)} \\
	\end{bmatrix}	
	\begin{bmatrix}
		b_{in}^{(i)} \\
		b_{in}^{(V)} \\
	\end{bmatrix},
\end{equation}
where the superscript $(V)$ denotes the evanescent electric potential wave in the vacuum gap.

The first boundary condition in Eqs.\eqref{eq:boundary_condition} can then be rearranged 
by moving all the outgoing (incoming) waves to the left (right) side, giving  
\begin{equation}\label{eq:smtx_boundarycondition}
		\bigg[
		\pmb{U}_1^{(1)},\pmb{U}_2^{(1)},\pmb{U}_3^{(1)},\pmb{U}_4^{(1)},-\pmb{U}_{V_+}
		\bigg]
		\begin{bmatrix}
			b_1^{(1)} \\
			b_2^{(1)} \\
			b_3^{(1)} \\
			b_4^{(1)} \\
			b_{V_+} \\
		\end{bmatrix}
		=
		\bigg[
		-\pmb{U}_{in}^{(1)},\pmb{U}_{V_-}
		\bigg]
		\begin{bmatrix}
			b_{in}^{(1)} \\
			b_{V_-} \\
		\end{bmatrix}. 
\end{equation}
The second boundary condition follows from Eq.\eqref{eq:smtx_boundarycondition} by changing the medium index, and exchanging the incoming and outgoing vacuum waves 
\begin{equation}\label{eq:smtx_boundarycondition2}
		\bigg[
		\pmb{U}_1^{(2)},\pmb{U}_2^{(2)},\pmb{U}_3^{(2)},\pmb{U}_4^{(2)},-\pmb{U}_{V_-}
		\bigg]
		\begin{bmatrix}
			b_1^{(2)} \\
			b_2^{(2)} \\
			b_3^{(2)} \\
			b_4^{(2)} \\
			b_{V_-} \\
		\end{bmatrix}
		=
		\bigg[
		-\pmb{U}_{in}^{(2)},\pmb{U}_{V_+}
		\bigg]
		\begin{bmatrix}
			b_{in}^{(2)} \\
			b_{V_+} \\
		\end{bmatrix}.
\end{equation}
By comparing Eqs.\eqref{eq:smtx_boundarycondition} and \eqref{eq:smtx_boundarycondition2} with Eq.\eqref{eq:smtx_form}, we obtain the expressions for the scattering matrices $\pmb{S}^{(1)}$ and $\pmb{S}^{(2)}$ as given in Eq.\eqref{eq:sct_matrix}. Note that even if we have formally used an input wave from medium (2) in the definition of $\pmb{S}^{(2)}$, due to the linearity of the problem it will not affect how an input wave from medium (1) is transmitted or reflected. In the actual computation of $\pmb{S}^{(2)}$ for the case of input wave from medium (1), $-\pmb{U}_{in}^{(2)}$ can be set arbitrarily, for example to zero.  

\section{Details of the analytical solution example}\label{apd:analyticaldetails}

In the analytical example we presented in Section \ref{sec:analytical}, a  hexagonal $6mm$ symmetry crystal was rotated in such way that its crystallographic $c$-axis is aligned with the solid-vacuum interface and is perpendicular to the sagittal (incident) plane. The material parameters of the rotated crystal,  $\pmb{\epsilon}^S$ the electric permittivity at constant strain, $\pmb{e}$ the piezoelectric stress, and $\pmb{c}^E$ the elastic stiffness at constant electric field,  can be obtained using the method provided in Appendix \ref{apd:crystal_rotation}. To be specific, the crystal is rotated about the $x$-axis by $90^\circ$ following the right-hand rule, after which the rotated material tensors read as:
\begin{align}
    \pmb{e} &= \left[\begin{matrix}0 & 0 & 0 & 0 & 0 & - e_{x5}\\- e_{z1} & - e_{z3} & - e_{z1} & 0 & 0 & 0\\0 & 0 & 0 & - e_{x5} & 0 & 0\end{matrix}\right] \\
    \pmb{\epsilon}^S &=\left[\begin{matrix}\epsilon_{xx} & 0 & 0\\0 & \epsilon_{zz} & 0\\0 & 0 & \epsilon_{xx}\end{matrix}\right] \\
    \pmb{c}^E &= \left[\begin{matrix}c_{11} & c_{13} & c_{12} & 0 & 0 & 0\\c_{13} & c_{33} & c_{13} & 0 & 0 & 0\\c_{12} & c_{13} & c_{11} & 0 & 0 & 0\\0 & 0 & 0 & c_{44} & 0 & 0\\0 & 0 & 0 & 0 & c_{66} & 0\\0 & 0 & 0 & 0 & 0 & c_{44}\end{matrix}\right]\ .
\end{align}

In Appendix \ref{apd:stroh_formalism}, the general approach for computing the extended Stroh $8\times8$ matrix $\pmb{N}$ was described. However, this matrix can be significantly simplified in the analytical example in Section \ref{sec:analytical}. This is because for this high symmetry case, piezoelectric response appears only along the crystal $c$-axis, which is aligned with the $y$-axis of the laboratory coordinates after the rotation (as shown in Fig.2(a)), and there is no mode conversion as stated in the main text. Therefore, only the $y$-axis components,  $u_y$ in displacement and $\sigma_{yz}$ in stress, enter the boundary conditions, Eqs.\eqref{eq:general_boundcond}, and thus a reduced $4\times4$ Stroh matrix $\pmb{N}$ and four-dimensional eigen-vectors $\pmb{\xi}=[u_y,\phi,L_{yz},D]^T$ are sufficient to solve the scattering problem at hand, involving only SH and E wave modes. 

The explicit expression of the reduced Stroh matrix $\pmb{N}$ is given in the main text in Eq.\eqref{eq:N4}, and its eigenvalues and Stroh-normalized eigenvectors are:
\begin{widetext}
\begin{align*}
p_1&=-i,\,&&
\pmb{\xi}_{1}=
\left[
0,
\sqrt{\frac{i}{2\epsilon_{xx}}},
\frac{e_{x5}\sqrt{-i}}{\sqrt{2\epsilon_{xx}}},
\frac{\sqrt{-i\epsilon_{xx}}}{\sqrt{2}}
\right]^T\\
p_2&=i,\,&&
\pmb{\xi}_{2}=
\left[
0,
\frac{\sqrt{-i}}{\sqrt{2\epsilon_{xx}}},
\frac{e_{x5}\sqrt{i}}{\sqrt{2\epsilon_{xx}}},
\frac{\sqrt{i\epsilon_{xx}}}{\sqrt{2}}
\right]^T \\
p_3&=-\cot\theta_i,\,&&
\pmb{\xi}_{3}=
\left[
\frac{\sqrt{2}}{2}\sqrt{\frac{k^2\tan\theta_i}{\rho\omega^2}},
-\frac{\sqrt{2}e_{x5}}{2\epsilon_{xx}}\sqrt{\frac{k^2\tan\theta_i}{\rho\omega^2}},
\frac{\sqrt{2}}{2}\sqrt{\frac{\rho\omega^2}{k^2\tan\theta_i}},
0
\right]^T \\
p_4&=\cot\theta_i,\,&&
\pmb{\xi}_{4}=
\left[
\frac{\sqrt{2}}{2}\sqrt{\frac{-k^2\tan\theta_i}{\rho\omega^2}},
-\frac{\sqrt{2}e_{x5}}{2\epsilon_{xx}}\sqrt{\frac{-k^2\tan\theta_i}{\rho\omega^2}},
\frac{\sqrt{2}}{2}\sqrt{\frac{-\rho\omega^2}{k^2\tan\theta_i}},
0
\right]^T .\\
\end{align*}
\end{widetext}
The normalization conditions for the above solutions are $2(u_yL_{yz}+\phi D)=1$, and the dispersion relation is $\rho\omega^2=(c_{44}+e_{x5}^2/\epsilon_{xx})k^2$. The first two solutions ($p_1,p_2$) correspond to the two inhomogeneous E waves, and the last two ($p_3,p_4$) to the propagating SH waves.

\section{Crystallographic orientation}\label{apd:crystal_rotation}
To solve the tunneling problem for an arbitrary crystal orientation, a method for transforming the material tensors from a standard crystallographic orientation to a specific arbitrary rotation needs to be provided. The tensors in question are the electric permittivity at constant strain, $\pmb{\epsilon}^S_0$, the piezoelectric stress, $\pmb{e}_0$, and elastic stiffness at constant electric field, $\pmb{c}^E_0$, where the subscript $0$ refers to the standard  crystallographic orientation.

\begin{figure}[h]
	\centering
	\includegraphics[width=0.8\linewidth]{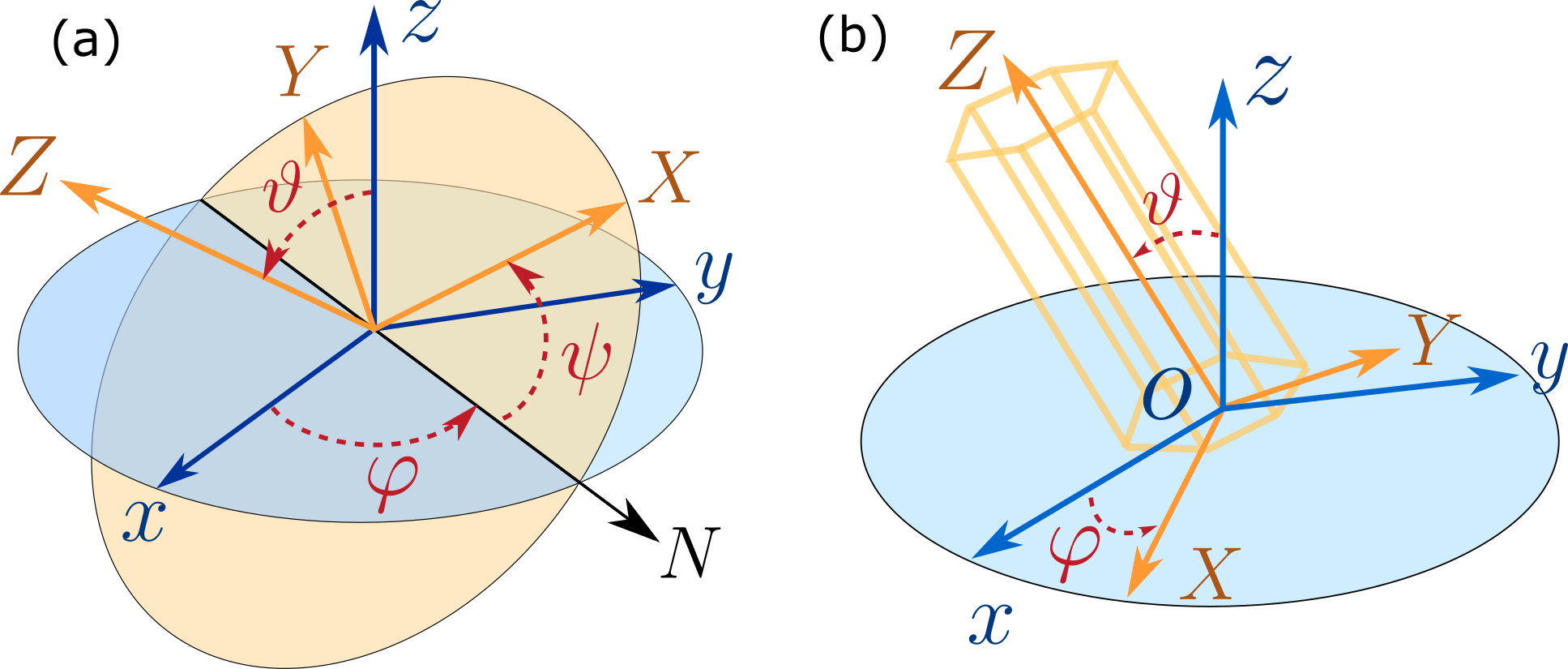}
	\caption{Demonstration of crystal rotation angles. (a) The general Euler angle system. (b) The cylindrical angle system for uniaxial crystals.}
	\label{fig:Eulerangles}
\end{figure}

To describe the orientation of a crystal with respect to a fixed laboratory coordinate system, we adopt the Euler angle system \cite{Goldstein}. In this system, we define two Cartesian frames $XYZ$ and $xyz$, the crystal intrinsic coordinates and the external fixed laboratory coordinates, respectively. The relation between these two frames can be fully expressed by three angles: $\vartheta$, $\varphi$ and $\psi$, as illustrated in Figure \ref{fig:Eulerangles}(a). 

Several different conventions of the sequence of elemental rotations can be used to acquire the material constants for a specific crystal orientation ($\vartheta$,$\varphi$,$\psi$). In this work, we adopted the widely used extrinsic $z$-$x$-$z$ rotation sequence, which rotates the crystal frame from initial overlap with the laboratory coordinates to the desired orientation. In this procedure, the crystal frame will first be rotated about the $z$-axis by an angle $\psi$ defined by the right-hand rule (counter-clockwise if viewed from top), followed by a second right-hand rotation of angle $\vartheta$ about the $x$-axis, and finally a third right-hand rotation of angle $\varphi$ about the $z$-axis.

The material constant tensors (represented by $\pmb{T}_{m\times n}$ matrices in the abbreviated index notation) can then be transformed to the rotated ones $\pmb{T}_{m\times n}'$ by using rotation transformation matrices $\pmb{R}$ \cite{Auld1973}:
\begin{equation}\label{eq:rotation_transformation}
	\pmb{T}_{m\times n}' = \pmb{R}_m\pmb{T}_{m\times n}\pmb{R}_n^T,
\end{equation}
where $\pmb{R}_m$ or $\pmb{R}_n$ are the two rotation transformation matrices required for a general $m \times n$ matrix. 
In our case, $\pmb{\epsilon}^S_0$ has $m,n = 3$, $\pmb{e}_0$ has $m = 3$ and $n=6$, and $\pmb{c}^E_0$ $m,n = 6$, so we need only two different dimensionalities of rotation matrices $\pmb{R}_3$ and $\pmb{R}_6$ for both $z-$ and $x$-axes, for a total of four rotation matrices. 
 
For the crystal rotations about the $x$- and $z$-axes by the right-hand rule angles $\xi_x$ and $\xi_z$, respectively, $\pmb{R}_3$ can be expressed \cite{Auld1973} as
\begin{align}
\label{eq:R3x}
	\pmb{R}_{3,x}(\xi_x) &=
	\begin{bmatrix}
		1 & 0 & 0\\
		0 & \cos{\xi_x } & -\sin{\xi_x }\\
		0 & \sin{\xi_x } & \cos{\xi_x }
	\end{bmatrix},
\\
\label{eq:R3z}
	\pmb{R}_{3,z}(\xi_z) &=
	\begin{bmatrix}
		\cos{\xi_z} & -\sin{\xi_z } & 0\\
		\sin{\xi_z } & \cos{\xi_z } & 0\\
		0 & 0 & 1
	\end{bmatrix}.
\end{align}
$\pmb{R}_6$, required for the higher rank $\pmb{e}$ and $\pmb{c}$ tensors, can be obtained from the Bond stress matrix \cite{Auld1973} as
\begin{widetext}
\begin{align}
\label{eq:R6x}
\pmb{R}_{6,x}(\xi_x)&=
\begin{bmatrix}
    1 & 0 & 0 & 0 & 0 & 0 \\
    0 & \cos^{2}{\left(\xi_x \right)} & \sin^{2}{\left(\xi_x \right)} & - 2 \sin{\left(\xi_x \right)} \cos{\left(\xi_x \right)} & 0 & 0\\
    0 & \sin^{2}{\left(\xi_x \right)} & \cos^{2}{\left(\xi_x \right)} & 2 \sin{\left(\xi_x \right)} \cos{\left(\xi_x \right)} & 0 & 0\\
    0 & \sin{\left(\xi_x \right)} \cos{\left(\xi_x \right)} & - \sin{\left(\xi_x \right)} \cos{\left(\xi_x \right)} & - \sin^{2}{\left(\xi_x \right)} + \cos^{2}{\left(\xi_x \right)} & 0 & 0\\
    0 & 0 & 0 & 0 & \cos{\left(\xi_x \right)} & \sin{\left(\xi_x \right)}\\
    0 & 0 & 0 & 0 & - \sin{\left(\xi_x \right)} & \cos{\left(\xi_x \right)}
\end{bmatrix}
\\
\label{eq:R6z}
\pmb{R}_{6,z}(\xi_z)&=
\begin{bmatrix}
    \cos^{2}{\left(\xi_z \right)} & \sin^{2}{\left(\xi_z \right)} & 0 & 0 & 0 & - 2 \sin{\left(\xi_z \right)} \cos{\left(\xi_z \right)}\\
    \sin^{2}{\left(\xi_z \right)} & \cos^{2}{\left(\xi_z \right)} & 0 & 0 & 0 & 2 \sin{\left(\xi_z \right)} \cos{\left(\xi_z \right)}\\
    0 & 0 & 1 & 0 & 0 & 0\\
    0 & 0 & 0 & \cos{\left(\xi_z \right)} & \sin{\left(\xi_z \right)} & 0\\
    0 & 0 & 0 & - \sin{\left(\xi_z \right)} & \cos{\left(\xi_z \right)} & 0\\
    \sin{\left(\xi_z \right)} \cos{\left(\xi_z \right)} & - \sin{\left(\xi_z \right)} \cos{\left(\xi_z \right)} & 0 & 0 & 0 & - \sin^{2}{\left(\xi_z \right)} + \cos^{2}{\left(\xi_z \right)}
\end{bmatrix}.
\end{align}
\end{widetext}

As a result, the material tensors are then obtained with the composite $z$-$x$-$z$ rotation as 
\begin{widetext}
\begin{equation}
	\begin{aligned}
		\pmb{\epsilon}^{S} & = 
		\pmb{R}_{3,z}(\varphi)
		\left\{
		\pmb{R}_{3,x}(\vartheta)
		\left[
		\pmb{R}_{3,z}(\psi)\pmb{\epsilon}^S_0\pmb{R}_{3,z}(\psi)^T
		\right]
		\pmb{R}_{3,x}(\vartheta)^T 
		\right\}
		\pmb{R}_{3,z}(\varphi)^T 
		\\
		\pmb{e} & = 		
		\pmb{R}_{3,z}(\varphi)
		\left\{
		\pmb{R}_{3,x}(\vartheta)
		\left[
		\pmb{R}_{3,z}(\psi)\pmb{e}_0\pmb{R}_{6,z}(\psi)^T
		\right]
		\pmb{R}_{6,x}(\vartheta)^T 
		\right\}
		\pmb{R}_{6,z}(\varphi)^T 
		\\
		\pmb{c}^{E} & = 
		\pmb{R}_{6,z}(\varphi)
		\left\{
		\pmb{R}_{6,x}(\vartheta)
		\left[
		\pmb{R}_{6,z}(\psi)\pmb{c}^E_0\pmb{R}_{6,z}(\psi)^T
		\right]
		\pmb{R}_{6,x}(\vartheta)^T 
		\right\}
		\pmb{R}_{6,z}(\varphi)^T. 
	\end{aligned}
\end{equation}
\end{widetext}
For a solid with uniaxial symmetry about the crystal $Z$-axis, such as in our example (the wurtzite hexagonal crystal ZnO), the first $z$-axis rotation will not change the material tensors. Thus for such a symmetry, the description of the orientation can be simplified from the Euler angle system to the cylindrical angle system, which uses only a zenith angle $\vartheta$ and an azimuthal angle $\varphi$, as shown in Figure.\ref{fig:Eulerangles}(b). The corresponding rotation transformations will also be reduced to a two-step procedure: first a right-hand rotation of $\vartheta$ about the $x$-axis, followed by a second rotation of $\varphi$ about the $z$-axis. The material tensors will then be obtained as
\begin{equation}
	\begin{aligned}
		\pmb{\epsilon}^{S} & = 
		\pmb{R}_{3,z}(\varphi)
		\left[
		\pmb{R}_{3,x}(\vartheta)\pmb{\epsilon}^S_0\pmb{R}_{3,x}(\vartheta)^T
		\right]
		\pmb{R}_{3,z}(\varphi)^T 
		\\
		\pmb{e} & = 		
		\pmb{R}_{3,z}(\varphi)
		\left[
		\pmb{R}_{3,x}(\vartheta)\pmb{e}_0\pmb{R}_{6,x}(\vartheta)^T
		\right]
		\pmb{R}_{6,z}(\varphi)^T 
		\\
		\pmb{c}^{E} & = 
		\pmb{R}_{6,z}(\varphi)
		\left[
		\pmb{R}_{6,x}(\vartheta)\pmb{c}^E_0\pmb{R}_{6,x}(\vartheta)^T
		\right]
		\pmb{R}_{6,z}(\varphi)^T 
	\end{aligned}.
\end{equation}


\section{Wave mode assignment}\label{apd:wavemode_assignment}
Conventionally, there are two different approaches that have been widely used to categorize the three bulk elastic wave mode solutions. The first approach considers the relation between the particle displacement vector (also commonly known as the polarization vector) and the propagation direction of the wave (wave vector): when an elastic wave has a polarization that is (mostly) parallel to the propagating direction, it is identified as a \textit{(quasi-)longitudinal} wave, or an L mode; if a transverse wave is polarized (mostly) inside the plane of incidence and (not purely) perpendicular to the propagation direction, it is a \textit{vertically polarized (quasi-)shear} wave or an SV mode; and if a transverse wave is (mostly) perpendicular to both the plane of incidence and the propagation direction, it is a \textit{horizontally polarized (quasi-)shear} or an SH mode. 
For anisotropic crystals, the quasi-prefixes mostly apply, as pure L, SV and SH polarizations appear only in certain high symmetry propagation directions \cite{Auld1973}. 

The second approach is to compare the phase velocities $v=\omega/k$ of the wave modes, and to designate the mode from the fastest to the slowest as \textit{(quasi-)longitudinal} wave (L), \textit{fast (quasi-)transverse} wave (FT) and \textit{slow (quasi-)transverse} wave (ST).

It should be noted here that the choice for the categorization of the wave modes is a conceptual definition based on exactly the same set of solutions of the constitutive equations, and, therefore the choice of the categorization won't affect the results of the formalism discussed in this article. However, for completeness and for the benefit of the discussion of topics such as mode conversions, we provide here a procedure that can be programmed to consistently identify the wave modes based on both categorization approaches in this work. 

A total of eight eigenvalues $p_\alpha$ and their associated eigenvectors $\pmb{\xi}_\alpha$ can be obtained by solving the eigenfunction Eq.\eqref{eq:stroh_eigenfunc}. In this section, we will examine these solutions with four different categorization methods:
\begin{enumerate}
	\item Homogeneous or inhomogeneous wave
	\item Transmitted or reflected wave
	\item The mode categorized as longitudinal (L), fast transverse (FT), slow transverse (ST), or electric potential (E)
	\item The mode categorized as longitudinal (L), vertically polarized shear (SV), horizontally polarized shear (SH), or electric potential (E)
\end{enumerate} 
The electric potential mode E is an inhomogeneous wave mode solution that appears in piezoelectric scattering problems (within the quasistatic approximation), describing a solution where the energy is mostly contained in the electric fields \cite{Auld1973,Every1992}.    

First, for a wave solution that has an eigenvalue $p_\alpha=p_\alpha'+ip_\alpha''$, we examine the imaginary part: If $p_\alpha''=0$ ($p_\alpha''\neq0$), the wave will be categorized as a homogeneous wave (an inhomogeneous wave).

Second, for an inhomogeneous wave, the scattering direction of the wave can be determined by the imaginary part of the eigenvalue: If $p_\alpha''>0$ ($p_\alpha''<0$) the wave will be categorized as a transmitted (reflected) wave. This follows from the principle that the physically allowed inhomogeneous wave solution can only decay (and not grow) from the interface. In contrast, for a homogeneous plane wave, the direction of the power flow should be examined, as the normal components of the wave vector and the power flow can have different directions in general. By acknowledging the time-averaged Poynting vector in Eq.\eqref{eq:Poynting_vector}, a wave with $P_{n\alpha}<0$ ($P_{n\alpha}>0$) is categorized as a transmitted (reflected) wave.

The aforementioned wave modes (e.g. FT, SV, etc.) are defined from a partial set of characteristics of the wave solutions, such as phase velocity, polarization vector, etc. Therefore, it can in some cases be tricky to fully map such simplified mode definitions to the corresponding full solutions, and ambiguity can arise. For example, in some cases four scattered bulk modes can be excited  simultaneously (without the excitation of the inhomogeneous E mode) due to a strong electromechanical coupling \cite{Every1992}. 
Therefore, it should be kept in mind that the mode categorization method presented here is not a fully robust and generally applicable algorithm.

To assign the modes within the set L, FT, ST and E, we first compare the magnitudes of the imaginary parts $|p_\alpha''|$ of all the inhomogeneous evanescent waves ($p_\alpha''\neq0$), and identify them based on the ordering $|p_E''|>|p_L''|>|p_{FT}''|\geq|p_{ST}''|$ (always starting from the E-mode, if fewer than four inhomogeneous modes exist). For the remaining unassigned homogeneous modes, the phase velocities $v_\alpha^2=v_x^2/(1+p_\alpha^2)$ will be examined, and the wave modes are assigned in the order $v_{L}^2>v_{FT}^2\geq v_{ST}^2$, starting from the first unassigned mode. This means that if for example the L-mode was identified already as inhomogeneous, the fastest homogeneous mode would then be FT.


Finally, if one wishes to to assign the modes within the set quasi- L, SV, SH and E, the polarization vectors $\pmb{A}_\alpha$ of the eigenvector solutions should be examined. However, we still first identify the inhomogeneous wave modes with the method described above, based on the magnitudes of the imaginary parts of the eigenvalues, as there are often no clear general differences between the eigenvectors of the surface (inhomogeneous) modes. 

In contrast, for the homogeneous modes, definitions based on the polarization vector exist. 
  We identify them by comparing  the polarization vector with the wave vector and the unit normal vector of the sagittal plane. If quasi-L mode is still available for assignment, it can be identified from $|\pmb{k}\cdot\pmb{A}_L|>|\pmb{k}\cdot\pmb{A}_{SV,SH}|$. Within the coordinate system of this article, quasi-SV and quasi-SH modes can be identified from the relation $|[0,1,0]^T\cdot\pmb{A}_{SV}|<|[0,1,0]^T\cdot\pmb{A}_{SH}|$. 


\section{Common cut planes for a hexagonal crystal}\label{apd:crystal_cut}
For hexagonal crystals, the four basis vector Miller-Bravais index system $\{hkil\}$ is commonly used to designate a crystallographic plane family \cite{schwartzenbach2003}. These indices can be related to the crystal rotations, described in Section \ref{apd:crystal_rotation}, by $\vartheta=\angle_{\{hkil\}}$, in which $\angle_{\{hkil\}}$ is the angle between the plane normal and the crystal $Z$-axis, and can be calculated from
\begin{align*}
	\angle_{\{hkil\}} &= \arccos\left(
	\frac{\overrightarrow{(hkl)}\cdot\overrightarrow{(001)}}{|\overrightarrow{(hkl)}||\overrightarrow{(001)}|} 
	\right) \\
	&= \arccos\left[
	al\left(
	\frac{4}{3}c^2(h^2+k^2+hk)+a^2l^2
	\right)^{-\frac{1}{2}}
	\right],
\end{align*}
where $a$ and $c$ are the in-plane (X,Y) and out-of-plane (Z) lattice constants of the crystal, respectively. The common crystallographic plane families for ZnO are given in Table \ref{tab:crystal_plane} with their corresponding $\angle_{\{hkil\}}$.

\setlength{\tabcolsep}{12pt}
\begin{ruledtabular}
\begin{table}[h]
	\centering
	\begin{tabular}{ccc}
		Plane name & Miller index & $\angle_{\{hkil\}}$ \\
		\hline
		a & $\{11\overline{2}0\}$ & $90^\circ$ \\
		m & $\{10\overline{1}0\}$ & $90^\circ$ \\
		c & $\{0001\}$ & $0^\circ$ \\
		r & $\{1\overline{1}02\}$ & $42.78^\circ$ \\
		n & $\{11\overline{2}3\}$ & $46.89^\circ$ \\
		s & $\{10\overline{1}1\}$ & $61.61^\circ$ 
	\end{tabular}
	\caption{The common cut planes of hexagonal crystals: plane names, Miller indices and rotation angles (ZnO).}
	\label{tab:crystal_plane}
\end{table}
\end{ruledtabular}



\begin{thebibliography}{51}%
\makeatletter
\providecommand \@ifxundefined [1]{%
 \@ifx{#1\undefined}
}%
\providecommand \@ifnum [1]{%
 \ifnum #1\expandafter \@firstoftwo
 \else \expandafter \@secondoftwo
 \fi
}%
\providecommand \@ifx [1]{%
 \ifx #1\expandafter \@firstoftwo
 \else \expandafter \@secondoftwo
 \fi
}%
\providecommand \natexlab [1]{#1}%
\providecommand \enquote  [1]{``#1''}%
\providecommand \bibnamefont  [1]{#1}%
\providecommand \bibfnamefont [1]{#1}%
\providecommand \citenamefont [1]{#1}%
\providecommand \href@noop [0]{\@secondoftwo}%
\providecommand \href [0]{\begingroup \@sanitize@url \@href}%
\providecommand \@href[1]{\@@startlink{#1}\@@href}%
\providecommand \@@href[1]{\endgroup#1\@@endlink}%
\providecommand \@sanitize@url [0]{\catcode `\\12\catcode `\$12\catcode
  `\&12\catcode `\#12\catcode `\^12\catcode `\_12\catcode `\%12\relax}%
\providecommand \@@startlink[1]{}%
\providecommand \@@endlink[0]{}%
\providecommand \url  [0]{\begingroup\@sanitize@url \@url }%
\providecommand \@url [1]{\endgroup\@href {#1}{\urlprefix }}%
\providecommand \urlprefix  [0]{URL }%
\providecommand \Eprint [0]{\href }%
\providecommand \doibase [0]{https://doi.org/}%
\providecommand \selectlanguage [0]{\@gobble}%
\providecommand \bibinfo  [0]{\@secondoftwo}%
\providecommand \bibfield  [0]{\@secondoftwo}%
\providecommand \translation [1]{[#1]}%
\providecommand \BibitemOpen [0]{}%
\providecommand \bibitemStop [0]{}%
\providecommand \bibitemNoStop [0]{.\EOS\space}%
\providecommand \EOS [0]{\spacefactor3000\relax}%
\providecommand \BibitemShut  [1]{\csname bibitem#1\endcsname}%
\let\auto@bib@innerbib\@empty
\bibitem [{\citenamefont {Royer}\ and\ \citenamefont
  {Dieulesant}(2000{\natexlab{a}})}]{RoyerII}%
  \BibitemOpen
  \bibfield  {author} {\bibinfo {author} {\bibfnamefont {D.}~\bibnamefont
  {Royer}}\ and\ \bibinfo {author} {\bibfnamefont {E.}~\bibnamefont
  {Dieulesant}},\ }\href@noop {} {\emph {\bibinfo {title} {{Elastic Waves in
  Solids II, Generation, acousto-optic interaction, applications}}}}\ (\bibinfo
   {publisher} {Springer-Verlag},\ \bibinfo {address} {Berlin Heidelberg},\
  \bibinfo {year} {2000})\BibitemShut {NoStop}%
\bibitem [{\citenamefont {Royer}\ and\ \citenamefont
  {Dieulesant}(2000{\natexlab{b}})}]{RoyerI}%
  \BibitemOpen
  \bibfield  {author} {\bibinfo {author} {\bibfnamefont {D.}~\bibnamefont
  {Royer}}\ and\ \bibinfo {author} {\bibfnamefont {E.}~\bibnamefont
  {Dieulesant}},\ }\href@noop {} {\emph {\bibinfo {title} {{Elastic Waves in
  Solids I, Free and guided propagation}}}}\ (\bibinfo  {publisher}
  {Springer-Verlag},\ \bibinfo {address} {Berlin Heidelberg},\ \bibinfo {year}
  {2000})\BibitemShut {NoStop}%
\bibitem [{\citenamefont {Auld}(1990)}]{Auld1973}%
  \BibitemOpen
  \bibfield  {author} {\bibinfo {author} {\bibfnamefont {B.~A.}\ \bibnamefont
  {Auld}},\ }\href@noop {} {\emph {\bibinfo {title} {{Acoustic Fields and Waves
  in Solids}}}},\ \bibinfo {edition} {2nd}\ ed.\ (\bibinfo  {publisher}
  {Krieger},\ \bibinfo {address} {Malabar, Florida},\ \bibinfo {year}
  {1990})\BibitemShut {NoStop}%
\bibitem [{\citenamefont {Budaev}\ and\ \citenamefont
  {Bogy}(2011)}]{Budaev2011}%
  \BibitemOpen
  \bibfield  {author} {\bibinfo {author} {\bibfnamefont {B.~V.}\ \bibnamefont
  {Budaev}}\ and\ \bibinfo {author} {\bibfnamefont {D.~B.}\ \bibnamefont
  {Bogy}},\ }\bibfield  {title} {\bibinfo {title} {On the role of acoustic
  waves (phonons) in equilibrium heat exchange across a vacuum gap},\ }\href
  {https://doi.org/10.1063/1.3623433} {\bibfield  {journal} {\bibinfo
  {journal} {Appl. Phys. Lett.}\ }\textbf {\bibinfo {volume} {99}},\ \bibinfo
  {pages} {053109} (\bibinfo {year} {2011})}\BibitemShut {NoStop}%
\bibitem [{\citenamefont {Ezzahri}\ and\ \citenamefont
  {Joulain}(2014)}]{Ezzahri2014}%
  \BibitemOpen
  \bibfield  {author} {\bibinfo {author} {\bibfnamefont {Y.}~\bibnamefont
  {Ezzahri}}\ and\ \bibinfo {author} {\bibfnamefont {K.}~\bibnamefont
  {Joulain}},\ }\bibfield  {title} {\bibinfo {title} {Vacuum-induced phonon
  transfer between two solid dielectric materials: Illustrating the case of
  casimir force coupling},\ }\href {https://doi.org/10.1103/PhysRevB.90.115433}
  {\bibfield  {journal} {\bibinfo  {journal} {Phys. Rev. B}\ }\textbf {\bibinfo
  {volume} {90}},\ \bibinfo {pages} {115433} (\bibinfo {year}
  {2014})}\BibitemShut {NoStop}%
\bibitem [{\citenamefont {Xiong}\ \emph {et~al.}(2014)\citenamefont {Xiong},
  \citenamefont {Yang}, \citenamefont {Kosevich}, \citenamefont {Chalopin},
  \citenamefont {D'Agosta}, \citenamefont {Cortona},\ and\ \citenamefont
  {Volz}}]{Xiong2014}%
  \BibitemOpen
  \bibfield  {author} {\bibinfo {author} {\bibfnamefont {S.}~\bibnamefont
  {Xiong}}, \bibinfo {author} {\bibfnamefont {K.}~\bibnamefont {Yang}},
  \bibinfo {author} {\bibfnamefont {Y.~A.}\ \bibnamefont {Kosevich}}, \bibinfo
  {author} {\bibfnamefont {Y.}~\bibnamefont {Chalopin}}, \bibinfo {author}
  {\bibfnamefont {R.}~\bibnamefont {D'Agosta}}, \bibinfo {author}
  {\bibfnamefont {P.}~\bibnamefont {Cortona}},\ and\ \bibinfo {author}
  {\bibfnamefont {S.}~\bibnamefont {Volz}},\ }\bibfield  {title} {\bibinfo
  {title} {Classical to quantum transition of heat transfer between two silica
  clusters},\ }\href {https://doi.org/10.1103/PhysRevLett.112.114301}
  {\bibfield  {journal} {\bibinfo  {journal} {Phys. Rev. Lett.}\ }\textbf
  {\bibinfo {volume} {112}},\ \bibinfo {pages} {114301} (\bibinfo {year}
  {2014})}\BibitemShut {NoStop}%
\bibitem [{\citenamefont {Chiloyan}\ \emph {et~al.}(2015)\citenamefont
  {Chiloyan}, \citenamefont {Garg}, \citenamefont {Esfarjani},\ and\
  \citenamefont {Chen}}]{Chiloyan2015}%
  \BibitemOpen
  \bibfield  {author} {\bibinfo {author} {\bibfnamefont {V.}~\bibnamefont
  {Chiloyan}}, \bibinfo {author} {\bibfnamefont {J.}~\bibnamefont {Garg}},
  \bibinfo {author} {\bibfnamefont {K.}~\bibnamefont {Esfarjani}},\ and\
  \bibinfo {author} {\bibfnamefont {G.}~\bibnamefont {Chen}},\ }\bibfield
  {title} {\bibinfo {title} {{Transition from near-field thermal radiation to
  phonon heat conduction at sub-nanometre gaps}},\ }\href
  {https://doi.org/10.1038/ncomms7755} {\bibfield  {journal} {\bibinfo
  {journal} {Nat. Comm.}\ }\textbf {\bibinfo {volume} {6}},\ \bibinfo {pages}
  {7755} (\bibinfo {year} {2015})}\BibitemShut {NoStop}%
\bibitem [{\citenamefont {Pendry}\ \emph {et~al.}(2016)\citenamefont {Pendry},
  \citenamefont {Sasihithlu},\ and\ \citenamefont {Craster}}]{Pendry2016}%
  \BibitemOpen
  \bibfield  {author} {\bibinfo {author} {\bibfnamefont {J.~B.}\ \bibnamefont
  {Pendry}}, \bibinfo {author} {\bibfnamefont {K.}~\bibnamefont {Sasihithlu}},\
  and\ \bibinfo {author} {\bibfnamefont {R.~V.}\ \bibnamefont {Craster}},\
  }\bibfield  {title} {\bibinfo {title} {Phonon-assisted heat transfer between
  vacuum-separated surfaces},\ }\href
  {https://doi.org/10.1103/PhysRevB.94.075414} {\bibfield  {journal} {\bibinfo
  {journal} {Phys. Rev. B}\ }\textbf {\bibinfo {volume} {94}},\ \bibinfo
  {pages} {075414} (\bibinfo {year} {2016})}\BibitemShut {NoStop}%
\bibitem [{\citenamefont {Volokitin}(2020)}]{Volokitin2020}%
  \BibitemOpen
  \bibfield  {author} {\bibinfo {author} {\bibfnamefont {A.~I.}\ \bibnamefont
  {Volokitin}},\ }\bibfield  {title} {\bibinfo {title} {Contribution of the
  acoustic waves to near-field heat transfer},\ }\href
  {https://doi.org/10.1088/1361-648x/ab71a5} {\bibfield  {journal} {\bibinfo
  {journal} {J. Phys.: Condens. Matter}\ }\textbf {\bibinfo {volume} {32}},\
  \bibinfo {pages} {215001} (\bibinfo {year} {2020})}\BibitemShut {NoStop}%
\bibitem [{\citenamefont {Balakirev}\ \emph {et~al.}(1978)\citenamefont
  {Balakirev}, \citenamefont {Bogdanov},\ and\ \citenamefont
  {Gorchakov}}]{Balakirev1978}%
  \BibitemOpen
  \bibfield  {author} {\bibinfo {author} {\bibfnamefont {M.}~\bibnamefont
  {Balakirev}}, \bibinfo {author} {\bibfnamefont {S.}~\bibnamefont
  {Bogdanov}},\ and\ \bibinfo {author} {\bibfnamefont {A.}~\bibnamefont
  {Gorchakov}},\ }\bibfield  {title} {\bibinfo {title} {{Tunneling of
  ultrasonic wave through a gap between lithium iodate crystals}},\ }\href@noop
  {} {\bibfield  {journal} {\bibinfo  {journal} {Fiz. Tverd. Tela (Leningrad)}\
  }\textbf {\bibinfo {volume} {20}},\ \bibinfo {pages} {587} (\bibinfo {year}
  {1978})},\ \bibinfo {note} {[Sov. Phys. Solid State, \textbf{20}, 338
  (1978)]}\BibitemShut {NoStop}%
\bibitem [{\citenamefont {Prunnila}\ and\ \citenamefont
  {Meltaus}(2010)}]{Prunnila2010}%
  \BibitemOpen
  \bibfield  {author} {\bibinfo {author} {\bibfnamefont {M.}~\bibnamefont
  {Prunnila}}\ and\ \bibinfo {author} {\bibfnamefont {J.}~\bibnamefont
  {Meltaus}},\ }\bibfield  {title} {\bibinfo {title} {{Acoustic phonon
  tunneling and heat transport due to evanescent electric fields}},\ }\href
  {https://doi.org/10.1103/PhysRevLett.105.125501} {\bibfield  {journal}
  {\bibinfo  {journal} {Phys. Rev. Lett.}\ }\textbf {\bibinfo {volume} {105}},\
  \bibinfo {pages} {125501} (\bibinfo {year} {2010})}\BibitemShut {NoStop}%
\bibitem [{\citenamefont {Kaliski}(1966)}]{Kaliski1966}%
  \BibitemOpen
  \bibfield  {author} {\bibinfo {author} {\bibfnamefont {S.}~\bibnamefont
  {Kaliski}},\ }\bibfield  {title} {\bibinfo {title} {{The passage of an
  ultrasonic wave across a contactless junction between two piezoelectric
  bodies}},\ }\href@noop {} {\bibfield  {journal} {\bibinfo  {journal} {Proc.
  Vibr. Probl. Warsaw}\ }\textbf {\bibinfo {volume} {7}},\ \bibinfo {pages}
  {95} (\bibinfo {year} {1966})}\BibitemShut {NoStop}%
\bibitem [{\citenamefont {Balakirev}\ and\ \citenamefont
  {Gorchakov}(1977)}]{Balakirev1977}%
  \BibitemOpen
  \bibfield  {author} {\bibinfo {author} {\bibfnamefont {M.}~\bibnamefont
  {Balakirev}}\ and\ \bibinfo {author} {\bibfnamefont {A.}~\bibnamefont
  {Gorchakov}},\ }\bibfield  {title} {\bibinfo {title} {{Leakage of an elastic
  wave across a gap between piezoelectrics}},\ }\href@noop {} {\bibfield
  {journal} {\bibinfo  {journal} {Fiz. Tverd. Tela (Leningrad)}\ }\textbf
  {\bibinfo {volume} {19}},\ \bibinfo {pages} {571} (\bibinfo {year} {1977})},\
  \bibinfo {note} {[Sov. Phys. Solid State, \textbf{19}, 327
  (1977)]}\BibitemShut {NoStop}%
\bibitem [{\citenamefont {Auld}(1981)}]{Auld1981}%
  \BibitemOpen
  \bibfield  {author} {\bibinfo {author} {\bibfnamefont {B.~A.}\ \bibnamefont
  {Auld}},\ }\bibfield  {title} {\bibinfo {title} {{Wave propagation and
  resonance in piezoelectric materials}},\ }\href
  {https://doi.org/10.1121/1.387223} {\bibfield  {journal} {\bibinfo  {journal}
  {J. Acoust. Soc. Am.}\ }\textbf {\bibinfo {volume} {70}},\ \bibinfo {pages}
  {1577} (\bibinfo {year} {1981})}\BibitemShut {NoStop}%
\bibitem [{\citenamefont {Barnett}\ and\ \citenamefont
  {Lothe}(1975)}]{Barnett1975}%
  \BibitemOpen
  \bibfield  {author} {\bibinfo {author} {\bibfnamefont {D.~M.}\ \bibnamefont
  {Barnett}}\ and\ \bibinfo {author} {\bibfnamefont {J.}~\bibnamefont
  {Lothe}},\ }\bibfield  {title} {\bibinfo {title} {{Dislocations and line
  charges in anisotropic piezoelectric insulators}},\ }\href
  {https://doi.org/10.1002/pssb.2220670108} {\bibfield  {journal} {\bibinfo
  {journal} {Phys. Status Solidi B}\ }\textbf {\bibinfo {volume} {67}},\
  \bibinfo {pages} {105} (\bibinfo {year} {1975})}\BibitemShut {NoStop}%
\bibitem [{\citenamefont {Lothe}\ and\ \citenamefont
  {Barnett}(1976)}]{Lothe1976}%
  \BibitemOpen
  \bibfield  {author} {\bibinfo {author} {\bibfnamefont {J.}~\bibnamefont
  {Lothe}}\ and\ \bibinfo {author} {\bibfnamefont {D.~M.}\ \bibnamefont
  {Barnett}},\ }\bibfield  {title} {\bibinfo {title} {{Integral formalism for
  surface waves in piezoelectric crystals. Existence considerations}},\ }\href
  {https://doi.org/10.1063/1.322895} {\bibfield  {journal} {\bibinfo  {journal}
  {J. Appl. Phys.}\ }\textbf {\bibinfo {volume} {47}},\ \bibinfo {pages} {1799}
  (\bibinfo {year} {1976})}\BibitemShut {NoStop}%
\bibitem [{\citenamefont {Stroh}(1962)}]{Stroh1962}%
  \BibitemOpen
  \bibfield  {author} {\bibinfo {author} {\bibfnamefont {A.~N.}\ \bibnamefont
  {Stroh}},\ }\bibfield  {title} {\bibinfo {title} {{Steady state problems in
  anisotropic elasticity}},\ }\href {https://doi.org/10.1002/sapm196241177}
  {\bibfield  {journal} {\bibinfo  {journal} {J. Math. and Phys.}\ }\textbf
  {\bibinfo {volume} {41}},\ \bibinfo {pages} {77} (\bibinfo {year}
  {1962})}\BibitemShut {NoStop}%
\bibitem [{\citenamefont {Chadwick}\ and\ \citenamefont
  {Smith}(1977)}]{Chadwick1977}%
  \BibitemOpen
  \bibfield  {author} {\bibinfo {author} {\bibfnamefont {P.}~\bibnamefont
  {Chadwick}}\ and\ \bibinfo {author} {\bibfnamefont {G.~D.}\ \bibnamefont
  {Smith}},\ }\bibfield  {title} {\bibinfo {title} {{Foundations of the Theory
  of Surface Waves in Anisotropic Elastic Materials}},\ }\href
  {https://doi.org/10.1016/S0065-2156(08)70223-0} {\bibfield  {journal}
  {\bibinfo  {journal} {Adv. Appl. Mech.}\ }\textbf {\bibinfo {volume} {17}},\
  \bibinfo {pages} {303} (\bibinfo {year} {1977})}\BibitemShut {NoStop}%
\bibitem [{\citenamefont {Ting}(1996)}]{Ting1996book}%
  \BibitemOpen
  \bibfield  {author} {\bibinfo {author} {\bibfnamefont {T.~C.~T.}\
  \bibnamefont {Ting}},\ }\href@noop {} {\emph {\bibinfo {title} {{Anisotropic
  Elasticity: Theory and Applications}}}}\ (\bibinfo  {publisher} {Oxford
  University Press},\ \bibinfo {address} {New York},\ \bibinfo {year}
  {1996})\BibitemShut {NoStop}%
\bibitem [{\citenamefont {Ting}(2000)}]{Ting2000}%
  \BibitemOpen
  \bibfield  {author} {\bibinfo {author} {\bibfnamefont {T.~C.~T.}\
  \bibnamefont {Ting}},\ }\bibfield  {title} {\bibinfo {title} {{Recent
  developments in anisotropic elasticity}},\ }\href
  {https://doi.org/10.1016/S0020-7683(99)00102-X} {\bibfield  {journal}
  {\bibinfo  {journal} {Int. J. Solids Struct.}\ }\textbf {\bibinfo {volume}
  {37}},\ \bibinfo {pages} {401} (\bibinfo {year} {2000})}\BibitemShut
  {NoStop}%
\bibitem [{\citenamefont {Al'shits}\ \emph {et~al.}(1989)\citenamefont
  {Al'shits}, \citenamefont {Darinskii},\ and\ \citenamefont
  {Shuvalov}}]{Alshits1989}%
  \BibitemOpen
  \bibfield  {author} {\bibinfo {author} {\bibfnamefont {V.}~\bibnamefont
  {Al'shits}}, \bibinfo {author} {\bibfnamefont {A.}~\bibnamefont
  {Darinskii}},\ and\ \bibinfo {author} {\bibfnamefont {A.}~\bibnamefont
  {Shuvalov}},\ }\bibfield  {title} {\bibinfo {title} {{Theory of reflection of
  acoustoelectric waves in a semiinfinite piezoelectric medium. I. Metallized
  surface}},\ }\href@noop {} {\bibfield  {journal} {\bibinfo  {journal}
  {Kristallografiya}\ }\textbf {\bibinfo {volume} {34}},\ \bibinfo {pages}
  {1340} (\bibinfo {year} {1989})},\ \bibinfo {note} {[Sov. Phys. Crystallogr.
  \textbf{34}, 808 (1989)]}\BibitemShut {NoStop}%
\bibitem [{\citenamefont {Al'shits}\ \emph {et~al.}(1990)\citenamefont
  {Al'shits}, \citenamefont {Darinskii},\ and\ \citenamefont
  {Shuvalov}}]{Alshits1990}%
  \BibitemOpen
  \bibfield  {author} {\bibinfo {author} {\bibfnamefont {V.}~\bibnamefont
  {Al'shits}}, \bibinfo {author} {\bibfnamefont {A.}~\bibnamefont
  {Darinskii}},\ and\ \bibinfo {author} {\bibfnamefont {A.}~\bibnamefont
  {Shuvalov}},\ }\bibfield  {title} {\bibinfo {title} {{Theory of reflection of
  acoustoelectric waves in a semiinfinite piezoelectric medium. II.
  Nonmetallized surface}},\ }\href@noop {} {\bibfield  {journal} {\bibinfo
  {journal} {Kristallografiya}\ }\textbf {\bibinfo {volume} {35}},\ \bibinfo
  {pages} {7} (\bibinfo {year} {1990})},\ \bibinfo {note} {[Sov. Phys.
  Crystallogr. \textbf{35}, 1 (1990)]}\BibitemShut {NoStop}%
\bibitem [{\citenamefont {Al'shits}\ \emph {et~al.}(1991)\citenamefont
  {Al'shits}, \citenamefont {Darinskii},\ and\ \citenamefont
  {Shuvalov}}]{Alshits1991}%
  \BibitemOpen
  \bibfield  {author} {\bibinfo {author} {\bibfnamefont {V.}~\bibnamefont
  {Al'shits}}, \bibinfo {author} {\bibfnamefont {A.}~\bibnamefont
  {Darinskii}},\ and\ \bibinfo {author} {\bibfnamefont {A.}~\bibnamefont
  {Shuvalov}},\ }\bibfield  {title} {\bibinfo {title} {{Theory of reflection of
  acoustoelectric waves in a semiinfinite piezoelectric medium. III. Resonance
  reflection in the neighborhood of a branch of outflowing waves}},\
  }\href@noop {} {\bibfield  {journal} {\bibinfo  {journal} {Kristallografiya}\
  }\textbf {\bibinfo {volume} {36}},\ \bibinfo {pages} {284} (\bibinfo {year}
  {1991})},\ \bibinfo {note} {[Sov. Phys. Crystallogr. \textbf{36}, 145
  (1991)]}\BibitemShut {NoStop}%
\bibitem [{\citenamefont {Chung}\ and\ \citenamefont {Ting}(1995)}]{Chung1995}%
  \BibitemOpen
  \bibfield  {author} {\bibinfo {author} {\bibfnamefont {M.~Y.}\ \bibnamefont
  {Chung}}\ and\ \bibinfo {author} {\bibfnamefont {T.~C.~T.}\ \bibnamefont
  {Ting}},\ }\bibfield  {title} {\bibinfo {title} {{Line force, charge, and
  dislocation in anisotropic piezoelectric composite wedges and spaces}},\
  }\href {https://doi.org/10.1115/1.2895948} {\bibfield  {journal} {\bibinfo
  {journal} {J. Appl. Mech.}\ }\textbf {\bibinfo {volume} {62}},\ \bibinfo
  {pages} {423} (\bibinfo {year} {1995})}\BibitemShut {NoStop}%
\bibitem [{\citenamefont {Akamatsu}\ and\ \citenamefont
  {Tanuma}(1997)}]{Akamatsu1997}%
  \BibitemOpen
  \bibfield  {author} {\bibinfo {author} {\bibfnamefont {M.}~\bibnamefont
  {Akamatsu}}\ and\ \bibinfo {author} {\bibfnamefont {K.}~\bibnamefont
  {Tanuma}},\ }\bibfield  {title} {\bibinfo {title} {{Green's function of
  anisotropic piezoelectricity}},\ }\href
  {https://doi.org/10.1098/rspa.1997.0027} {\bibfield  {journal} {\bibinfo
  {journal} {Proc. R. Soc. London A}\ }\textbf {\bibinfo {volume} {453}},\
  \bibinfo {pages} {473} (\bibinfo {year} {1997})}\BibitemShut {NoStop}%
\bibitem [{\citenamefont {Hwu}(2008)}]{Hwu2008}%
  \BibitemOpen
  \bibfield  {author} {\bibinfo {author} {\bibfnamefont {C.}~\bibnamefont
  {Hwu}},\ }\bibfield  {title} {\bibinfo {title} {{Some explicit expressions of
  extended Stroh formalism for two-dimensional piezoelectric anisotropic
  elasticity}},\ }\href {https://doi.org/10.1016/j.ijsolstr.2008.03.025}
  {\bibfield  {journal} {\bibinfo  {journal} {Int. J. Solids Struct.}\ }\textbf
  {\bibinfo {volume} {45}},\ \bibinfo {pages} {4460} (\bibinfo {year}
  {2008})}\BibitemShut {NoStop}%
\bibitem [{\citenamefont {Lyubimov}\ \emph {et~al.}(1980)\citenamefont
  {Lyubimov}, \citenamefont {Alshits},\ and\ \citenamefont
  {Lothe}}]{Lyubimov1980}%
  \BibitemOpen
  \bibfield  {author} {\bibinfo {author} {\bibfnamefont {V.~N.}\ \bibnamefont
  {Lyubimov}}, \bibinfo {author} {\bibfnamefont {V.~I.}\ \bibnamefont
  {Alshits}},\ and\ \bibinfo {author} {\bibfnamefont {J.}~\bibnamefont
  {Lothe}},\ }\bibfield  {title} {\bibinfo {title} {{Body waves and quasi-body
  surface waves in a semi-infinite piezoelectric medium}},\ }\href@noop {}
  {\bibfield  {journal} {\bibinfo  {journal} {Kristallografiya}\ }\textbf
  {\bibinfo {volume} {25}},\ \bibinfo {pages} {33} (\bibinfo {year} {1980})},\
  \bibinfo {note} {[Sov. Phys. Crystallogr. \textbf{25}, 16
  (1980)]}\BibitemShut {NoStop}%
\bibitem [{\citenamefont {Darinskii}\ and\ \citenamefont
  {Weihnacht}(2003)}]{Darinskii2003}%
  \BibitemOpen
  \bibfield  {author} {\bibinfo {author} {\bibfnamefont {A.~N.}\ \bibnamefont
  {Darinskii}}\ and\ \bibinfo {author} {\bibfnamefont {M.}~\bibnamefont
  {Weihnacht}},\ }\bibfield  {title} {\bibinfo {title} {Quasi-bulk surface and
  leaky waves in piezoelectrics of unrestricted symmetry},\ }\href
  {https://doi.org/10.1098/rspa.2003.1142} {\bibfield  {journal} {\bibinfo
  {journal} {Proc. R. Soc. London A}\ }\textbf {\bibinfo {volume} {459}},\
  \bibinfo {pages} {2977} (\bibinfo {year} {2003})}\BibitemShut {NoStop}%
\bibitem [{\citenamefont {Al'shits}\ \emph {et~al.}(1993)\citenamefont
  {Al'shits}, \citenamefont {Darinskii},\ and\ \citenamefont
  {Shuvalov}}]{ALSHITS1993}%
  \BibitemOpen
  \bibfield  {author} {\bibinfo {author} {\bibfnamefont {V.~I.}\ \bibnamefont
  {Al'shits}}, \bibinfo {author} {\bibfnamefont {A.~N.}\ \bibnamefont
  {Darinskii}},\ and\ \bibinfo {author} {\bibfnamefont {A.~L.}\ \bibnamefont
  {Shuvalov}},\ }\bibfield  {title} {\bibinfo {title} {{Acoustoelectric waves
  in bicrystal media in conditions of a rigid contact or a vacuum gap at an
  interface}},\ }\href@noop {} {\bibfield  {journal} {\bibinfo  {journal}
  {Kristallografiya}\ }\textbf {\bibinfo {volume} {38}},\ \bibinfo {pages} {22}
  (\bibinfo {year} {1993})},\ \bibinfo {note} {[Crystallogr. Rep. \textbf{38},
  147 (1993)]}\BibitemShut {NoStop}%
\bibitem [{\citenamefont {Al'shits}\ \emph {et~al.}(1994)\citenamefont
  {Al'shits}, \citenamefont {Barnett}, \citenamefont {Darinskii},\ and\
  \citenamefont {Lothe}}]{Alshits1994}%
  \BibitemOpen
  \bibfield  {author} {\bibinfo {author} {\bibfnamefont {V.~I.}\ \bibnamefont
  {Al'shits}}, \bibinfo {author} {\bibfnamefont {D.~M.}\ \bibnamefont
  {Barnett}}, \bibinfo {author} {\bibfnamefont {A.~N.}\ \bibnamefont
  {Darinskii}},\ and\ \bibinfo {author} {\bibfnamefont {J.}~\bibnamefont
  {Lothe}},\ }\bibfield  {title} {\bibinfo {title} {{On the existence problem
  for localized acoustic waves on the interface between two piezocrystals}},\
  }\href {https://doi.org/10.1016/0165-2125(94)90049-3} {\bibfield  {journal}
  {\bibinfo  {journal} {Wave Motion}\ }\textbf {\bibinfo {volume} {20}},\
  \bibinfo {pages} {233} (\bibinfo {year} {1994})}\BibitemShut {NoStop}%
\bibitem [{\citenamefont {Darinskii}\ and\ \citenamefont
  {Weihnacht}(2006)}]{Darinskii2006}%
  \BibitemOpen
  \bibfield  {author} {\bibinfo {author} {\bibfnamefont {A.~N.}\ \bibnamefont
  {Darinskii}}\ and\ \bibinfo {author} {\bibfnamefont {M.}~\bibnamefont
  {Weihnacht}},\ }\bibfield  {title} {\bibinfo {title} {{Gap Acousto-Electric
  Waves in Structures of Arbitrary Anisotropy}},\ }\href
  {https://doi.org/10.1109/TUFFC.2006.1593380} {\bibfield  {journal} {\bibinfo
  {journal} {IEEE Trans. Ultrason. Ferroelectr. Freq. Control}\ }\textbf
  {\bibinfo {volume} {53}},\ \bibinfo {pages} {412} (\bibinfo {year}
  {2006})}\BibitemShut {NoStop}%
\bibitem [{\citenamefont {Gulyaev}\ and\ \citenamefont
  {Plessky}(1976)}]{Gulyaev1976}%
  \BibitemOpen
  \bibfield  {author} {\bibinfo {author} {\bibfnamefont {Y.~V.}\ \bibnamefont
  {Gulyaev}}\ and\ \bibinfo {author} {\bibfnamefont {V.~P.}\ \bibnamefont
  {Plessky}},\ }\bibfield  {title} {\bibinfo {title} {{Shear surface acoustic
  waves in dielectrics in the presence of an electric field}},\ }\href
  {https://doi.org/10.1016/0375-9601(76)90741-6} {\bibfield  {journal}
  {\bibinfo  {journal} {Phys. Lett. A}\ }\textbf {\bibinfo {volume} {56}},\
  \bibinfo {pages} {491} (\bibinfo {year} {1976})}\BibitemShut {NoStop}%
\bibitem [{\citenamefont {Gulyaev}\ and\ \citenamefont
  {Plesskii}(1977)}]{Gulyaev1977}%
  \BibitemOpen
  \bibfield  {author} {\bibinfo {author} {\bibfnamefont {Y.~V.}\ \bibnamefont
  {Gulyaev}}\ and\ \bibinfo {author} {\bibfnamefont {V.~P.}\ \bibnamefont
  {Plesskii}},\ }\bibfield  {title} {\bibinfo {title} {{Acoustic gap waves in
  piezoelectric materials}},\ }\href@noop {} {\bibfield  {journal} {\bibinfo
  {journal} {Akust. Zh.}\ }\textbf {\bibinfo {volume} {23}},\ \bibinfo {pages}
  {716} (\bibinfo {year} {1977})},\ \bibinfo {note} {[Sov. Phys. Acoustics
  \textbf{23}, 410 (1977)]}\BibitemShut {NoStop}%
\bibitem [{\citenamefont {Pak}(1992)}]{Pak1992}%
  \BibitemOpen
  \bibfield  {author} {\bibinfo {author} {\bibfnamefont {Y.~E.}\ \bibnamefont
  {Pak}},\ }\bibfield  {title} {\bibinfo {title} {{Linear electro-elastic
  fracture mechanics of piezoelectric materials}},\ }\href
  {https://doi.org/10.1007/BF00040857} {\bibfield  {journal} {\bibinfo
  {journal} {Int. J. Fract.}\ }\textbf {\bibinfo {volume} {54}},\ \bibinfo
  {pages} {79} (\bibinfo {year} {1992})}\BibitemShut {NoStop}%
\bibitem [{\citenamefont {Liang}\ \emph {et~al.}(1995)\citenamefont {Liang},
  \citenamefont {Han}, \citenamefont {Wang},\ and\ \citenamefont
  {Du}}]{Liang1995}%
  \BibitemOpen
  \bibfield  {author} {\bibinfo {author} {\bibfnamefont {J.}~\bibnamefont
  {Liang}}, \bibinfo {author} {\bibfnamefont {J.}~\bibnamefont {Han}}, \bibinfo
  {author} {\bibfnamefont {B.}~\bibnamefont {Wang}},\ and\ \bibinfo {author}
  {\bibfnamefont {S.}~\bibnamefont {Du}},\ }\bibfield  {title} {\bibinfo
  {title} {{Electroelastic modelling of anisotropic piezoelectric materials
  with an elliptic inclusion}},\ }\href
  {https://doi.org/10.1016/0020-7683(94)00299-C} {\bibfield  {journal}
  {\bibinfo  {journal} {Int. J. Solids and Struct.}\ }\textbf {\bibinfo
  {volume} {32}},\ \bibinfo {pages} {2989} (\bibinfo {year}
  {1995})}\BibitemShut {NoStop}%
\bibitem [{\citenamefont {Lu}\ \emph {et~al.}(2006)\citenamefont {Lu},
  \citenamefont {Lee},\ and\ \citenamefont {Lu}}]{Lu2006}%
  \BibitemOpen
  \bibfield  {author} {\bibinfo {author} {\bibfnamefont {P.}~\bibnamefont
  {Lu}}, \bibinfo {author} {\bibfnamefont {H.~P.}\ \bibnamefont {Lee}},\ and\
  \bibinfo {author} {\bibfnamefont {C.}~\bibnamefont {Lu}},\ }\bibfield
  {title} {\bibinfo {title} {{Exact solutions for simply supported functionally
  graded piezoelectric laminates by Stroh-like formalism}},\ }\href
  {https://doi.org/10.1016/j.compstruct.2005.01.012} {\bibfield  {journal}
  {\bibinfo  {journal} {Compos. Struct.}\ }\textbf {\bibinfo {volume} {72}},\
  \bibinfo {pages} {352} (\bibinfo {year} {2006})}\BibitemShut {NoStop}%
\bibitem [{\citenamefont {Darinskii}\ and\ \citenamefont
  {Shuvalov}(2019)}]{Darinskii2019}%
  \BibitemOpen
  \bibfield  {author} {\bibinfo {author} {\bibfnamefont {A.~N.}\ \bibnamefont
  {Darinskii}}\ and\ \bibinfo {author} {\bibfnamefont {A.~L.}\ \bibnamefont
  {Shuvalov}},\ }\bibfield  {title} {\bibinfo {title} {{Existence of surface
  acoustic waves in one-dimensional piezoelectric phononic crystals of general
  anisotropy}},\ }\href {https://doi.org/10.1103/PhysRevB.99.174305} {\bibfield
   {journal} {\bibinfo  {journal} {Phys. Rev. B}\ }\textbf {\bibinfo {volume}
  {99}},\ \bibinfo {pages} {174305} (\bibinfo {year} {2019})}\BibitemShut
  {NoStop}%
\bibitem [{\citenamefont {Benchabane}\ \emph {et~al.}(2006)\citenamefont
  {Benchabane}, \citenamefont {Khelif}, \citenamefont {Rauch}, \citenamefont
  {Robert},\ and\ \citenamefont {Laude}}]{Benchabane2006}%
  \BibitemOpen
  \bibfield  {author} {\bibinfo {author} {\bibfnamefont {S.}~\bibnamefont
  {Benchabane}}, \bibinfo {author} {\bibfnamefont {A.}~\bibnamefont {Khelif}},
  \bibinfo {author} {\bibfnamefont {J.~Y.}\ \bibnamefont {Rauch}}, \bibinfo
  {author} {\bibfnamefont {L.}~\bibnamefont {Robert}},\ and\ \bibinfo {author}
  {\bibfnamefont {V.}~\bibnamefont {Laude}},\ }\bibfield  {title} {\bibinfo
  {title} {{Evidence for complete surface wave band gap in a piezoelectric
  phononic crystal}},\ }\href {https://doi.org/10.1103/PhysRevE.73.065601}
  {\bibfield  {journal} {\bibinfo  {journal} {Phys. Rev. E}\ }\textbf {\bibinfo
  {volume} {73}},\ \bibinfo {pages} {065601(R)} (\bibinfo {year}
  {2006})}\BibitemShut {NoStop}%
\bibitem [{\citenamefont {Darinskii}(1997)}]{Darinskii1997}%
  \BibitemOpen
  \bibfield  {author} {\bibinfo {author} {\bibfnamefont {A.~N.}\ \bibnamefont
  {Darinskii}},\ }\bibfield  {title} {\bibinfo {title} {{On the theory of leaky
  waves in crystals}},\ }\href {https://doi.org/10.1016/s0165-2125(96)00031-5}
  {\bibfield  {journal} {\bibinfo  {journal} {Wave Motion}\ }\textbf {\bibinfo
  {volume} {25}},\ \bibinfo {pages} {35} (\bibinfo {year} {1997})}\BibitemShut
  {NoStop}%
\bibitem [{\citenamefont {Darinskii}(1998)}]{Darinskii1998b}%
  \BibitemOpen
  \bibfield  {author} {\bibinfo {author} {\bibfnamefont {A.~N.}\ \bibnamefont
  {Darinskii}},\ }\bibfield  {title} {\bibinfo {title} {{Leaky waves and the
  elastic wave resonance reflection on a crystal-thin solid layer interface.
  II. Leaky waves given rise to by exceptional bulk waves}},\ }\href
  {https://doi.org/10.1121/1.421052} {\bibfield  {journal} {\bibinfo  {journal}
  {J. Acoust. Soc. Am.}\ }\textbf {\bibinfo {volume} {103}},\ \bibinfo {pages}
  {1845} (\bibinfo {year} {1998})}\BibitemShut {NoStop}%
\bibitem [{\citenamefont {Polder}\ and\ \citenamefont {{Van
  Hove}}(1971)}]{Polder1971}%
  \BibitemOpen
  \bibfield  {author} {\bibinfo {author} {\bibfnamefont {D.}~\bibnamefont
  {Polder}}\ and\ \bibinfo {author} {\bibfnamefont {M.}~\bibnamefont {{Van
  Hove}}},\ }\bibfield  {title} {\bibinfo {title} {{Theory of Radiative Heat
  Transfer between Closely Spaced Bodies}},\ }\href
  {https://doi.org/10.1103/PhysRevB.4.3303} {\bibfield  {journal} {\bibinfo
  {journal} {Phys. Rev. B}\ }\textbf {\bibinfo {volume} {4}},\ \bibinfo {pages}
  {3303} (\bibinfo {year} {1971})}\BibitemShut {NoStop}%
\bibitem [{\citenamefont {Pendry}(1999)}]{Pendry1999}%
  \BibitemOpen
  \bibfield  {author} {\bibinfo {author} {\bibfnamefont {J.~B.}\ \bibnamefont
  {Pendry}},\ }\bibfield  {title} {\bibinfo {title} {{Radiative exchange of
  heat between nanostructures}},\ }\href
  {https://doi.org/10.1088/0953-8984/11/35/301} {\bibfield  {journal} {\bibinfo
   {journal} {J. Phys.: Condens. Matter}\ }\textbf {\bibinfo {volume} {11}},\
  \bibinfo {pages} {6621} (\bibinfo {year} {1999})}\BibitemShut {NoStop}%
\bibitem [{\citenamefont {Joulain}\ \emph {et~al.}(2005)\citenamefont
  {Joulain}, \citenamefont {Mulet}, \citenamefont {Marquier}, \citenamefont
  {Carminati},\ and\ \citenamefont {Greffet}}]{Joulain2005}%
  \BibitemOpen
  \bibfield  {author} {\bibinfo {author} {\bibfnamefont {K.}~\bibnamefont
  {Joulain}}, \bibinfo {author} {\bibfnamefont {J.-P.}\ \bibnamefont {Mulet}},
  \bibinfo {author} {\bibfnamefont {F.}~\bibnamefont {Marquier}}, \bibinfo
  {author} {\bibfnamefont {R.}~\bibnamefont {Carminati}},\ and\ \bibinfo
  {author} {\bibfnamefont {J.-J.}\ \bibnamefont {Greffet}},\ }\bibfield
  {title} {\bibinfo {title} {{Surface electromagnetic waves thermally excited:
  Radiative heat transfer, coherence properties and Casimir forces revisited in
  the near field}},\ }\href {https://doi.org/10.1016/j.surfrep.2004.12.002}
  {\bibfield  {journal} {\bibinfo  {journal} {Surf. Sci. Rep.}\ }\textbf
  {\bibinfo {volume} {57}},\ \bibinfo {pages} {59} (\bibinfo {year} {2005})},\
  \Eprint {https://arxiv.org/abs/0504068} {0504068} \BibitemShut {NoStop}%
\bibitem [{Note1()}]{Note1}%
  \BibitemOpen
  \bibinfo {note} {The sagittal plane has a rotational degree of freedom with
  respect to the normal of the interface plane (azimuth angle), which is
  equivalent to the rotation of the crystal azimuth angle $\varphi $. For the
  sake of simplicity and to avoid the duplication of the effect of this degree
  of freedom, we unambiguously take into account the azimuth angle by the
  rotation of the crystal (see Appendix \ref
  {apd:crystal_rotation}).}\BibitemShut {Stop}%
\bibitem [{\citenamefont {Mal{\'{e}}n}\ and\ \citenamefont
  {Lothe}(1970)}]{Malen1970}%
  \BibitemOpen
  \bibfield  {author} {\bibinfo {author} {\bibfnamefont {K.}~\bibnamefont
  {Mal{\'{e}}n}}\ and\ \bibinfo {author} {\bibfnamefont {J.}~\bibnamefont
  {Lothe}},\ }\bibfield  {title} {\bibinfo {title} {{Explicit Expressions for
  Dislocation Derivatives}},\ }\href {https://doi.org/10.1002/pssb.19700390130}
  {\bibfield  {journal} {\bibinfo  {journal} {Phys. Status Solidi B}\ }\textbf
  {\bibinfo {volume} {39}},\ \bibinfo {pages} {287} (\bibinfo {year}
  {1970})}\BibitemShut {NoStop}%
\bibitem [{Note2()}]{Note2}%
  \BibitemOpen
  \bibinfo {note} {The above results are strictly true only if $\protect \pmb
  {N}$ is non-degenerate (has a non-zero determinant). Slightly modified
  eigenvectors and normalization conditions have been determined in the
  opposite case \cite {Darinskii2003} taking place at the exact conditions for
  a critical angle (transonic state), where the reflected bulk wave carries
  energy only along the interface. Our discussion is meant for the general case
  to facilitate numerical computation, thus these conditions are special cases
  that do not have to be considered here, as numerical computation can be done
  very close to the exact conditions.}\BibitemShut {Stop}%
\bibitem [{\citenamefont {Court}\ and\ \citenamefont {von
  Willisen}(1964)}]{Court64}%
  \BibitemOpen
  \bibfield  {author} {\bibinfo {author} {\bibfnamefont {I.~N.}\ \bibnamefont
  {Court}}\ and\ \bibinfo {author} {\bibfnamefont {F.~K.}\ \bibnamefont {von
  Willisen}},\ }\bibfield  {title} {\bibinfo {title} {{Frustrated total
  internal reflection and application of its principle to laser cavity
  design}},\ }\href {https://doi.org/10.1364/AO.3.000719} {\bibfield  {journal}
  {\bibinfo  {journal} {Appl. Optics}\ }\textbf {\bibinfo {volume} {3}},\
  \bibinfo {pages} {719} (\bibinfo {year} {1964})}\BibitemShut {NoStop}%
\bibitem [{\citenamefont {Born}\ and\ \citenamefont {Wolf}(1999)}]{BornWolf}%
  \BibitemOpen
  \bibfield  {author} {\bibinfo {author} {\bibfnamefont {M.}~\bibnamefont
  {Born}}\ and\ \bibinfo {author} {\bibfnamefont {E.}~\bibnamefont {Wolf}},\
  }\href@noop {} {\emph {\bibinfo {title} {{Principles of Optics}}}},\ \bibinfo
  {edition} {7th}\ ed.\ (\bibinfo  {publisher} {Cambridge University Press},\
  \bibinfo {address} {Cambridge},\ \bibinfo {year} {1999})\BibitemShut
  {NoStop}%
\bibitem [{\citenamefont {Goldstein}(1980)}]{Goldstein}%
  \BibitemOpen
  \bibfield  {author} {\bibinfo {author} {\bibfnamefont {H.}~\bibnamefont
  {Goldstein}},\ }\href@noop {} {\emph {\bibinfo {title} {{Classical
  Mechanics}}}},\ \bibinfo {edition} {2nd}\ ed.\ (\bibinfo  {publisher}
  {Addison-Wesley Publishing, Reading, MA},\ \bibinfo {year}
  {1980})\BibitemShut {NoStop}%
\bibitem [{\citenamefont {Every}\ and\ \citenamefont
  {Neiman}(1992)}]{Every1992}%
  \BibitemOpen
  \bibfield  {author} {\bibinfo {author} {\bibfnamefont {A.~G.}\ \bibnamefont
  {Every}}\ and\ \bibinfo {author} {\bibfnamefont {V.~I.}\ \bibnamefont
  {Neiman}},\ }\bibfield  {title} {\bibinfo {title} {{Reflection of
  electroacoustic waves in piezoelectric solids: Mode conversion into four bulk
  waves}},\ }\href {https://doi.org/10.1063/1.350457} {\bibfield  {journal}
  {\bibinfo  {journal} {J. Appl. Phys.}\ }\textbf {\bibinfo {volume} {71}},\
  \bibinfo {pages} {6018} (\bibinfo {year} {1992})}\BibitemShut {NoStop}%
\bibitem [{\citenamefont {Schwarzenbach}(2003)}]{schwartzenbach2003}%
  \BibitemOpen
  \bibfield  {author} {\bibinfo {author} {\bibfnamefont {D.}~\bibnamefont
  {Schwarzenbach}},\ }\bibfield  {title} {\bibinfo {title} {{Note on
  Bravais–Miller indices}},\ }\href
  {https://doi.org/10.1107/S0021889803014778} {\bibfield  {journal} {\bibinfo
  {journal} {J. Appl. Cryst.}\ }\textbf {\bibinfo {volume} {36}},\ \bibinfo
  {pages} {1270} (\bibinfo {year} {2003})}\BibitemShut {NoStop}%
\end{thebibliography}
%

\end{document}